\documentclass[aps,prl,twocolumn,superscriptaddress,amsmath,amssymb,floatfix,preprintnumbers,nofootinbib,longbibliography]{revtex4-1}

\usepackage{graphicx} % Include figure files
\usepackage{dcolumn} % Align table columns on decimal point
\usepackage{bm} % bold math
\usepackage{hyperref} % add hypertext capabilities
\usepackage[mathlines]{lineno} % Enable numbering of text and display math
\usepackage{orcidlink}
\usepackage{soul}
\usepackage{placeins}
\usepackage[export]{adjustbox}
\usepackage{ amssymb }
\usepackage{rotating}
\usepackage{siunitx}
\usepackage{lineno}
\usepackage{tabularx}
\usepackage[utf8]{inputenc}
\usepackage[T1]{fontenc}

\hypersetup{
    pdfnewwindow=true,    % links in new window
    colorlinks=true,      % false: boxed links; true: colored links
    linkcolor=blue,       % color of internal links
    citecolor=blue,       % color of links to bibliography
    filecolor=blue,       % color of file links
    urlcolor=blue         % color of external links
}

\begin{document}
%TC:ignore
\title{The TeV Sun Rises: Discovery of Gamma rays from the Quiescent Sun with HAWC}

\author{A.~Albert}
\affiliation{Physics Division, Los Alamos National Laboratory, Los Alamos, NM 87544, USA }
\author{R.~Alfaro}
\affiliation{Instituto de F'{i}sica, Universidad Nacional Autónoma de México, Ciudad de Mexico, Mexico }

\author{C.~Alvarez}
\affiliation{Universidad Autónoma de Chiapas, Tuxtla Gutiérrez, Chiapas, México}

\author{J.C.~Arteaga-Velázquez}
\affiliation{Universidad Michoacana de San Nicolás de Hidalgo, Morelia, Mexico }

\author{D.~Avila Rojas}
\affiliation{Instituto de F'{i}sica, Universidad Nacional Autónoma de México, Ciudad de Mexico, Mexico }

\author{H.A.~Ayala Solares}
\affiliation{Department of Physics, Pennsylvania State University, University Park, PA, USA }

\author{R.~Babu}
\affiliation{Department of Physics, Michigan Technological University, Houghton, MI, USA }

\author{E.~Belmont-Moreno}
\affiliation{Instituto de F'{i}sica, Universidad Nacional Autónoma de México, Ciudad de Mexico, Mexico }

\author{C.~Brisbois}
\affiliation{Department of Physics, University of Maryland, College Park, MD, USA }

\author{K.S.~Caballero-Mora}
\affiliation{Universidad Autónoma de Chiapas, Tuxtla Gutiérrez, Chiapas, México}

\author{T.~Capistrán}
\affiliation{Instituto de Astronom'{i}a, Universidad Nacional Autónoma de México, Ciudad de Mexico, Mexico }

\author{A.~Carramiñana}
\affiliation{Instituto Nacional de Astrof'{i}sica, Óptica y Electrónica, Puebla, Mexico }

\author{S.~Casanova}
\affiliation{Instytut Fizyki Jadrowej im Henryka Niewodniczanskiego Polskiej Akademii Nauk, IFJ-PAN, Krakow, Poland }

\author{O.~Chaparro-Amaro}
\affiliation{Centro de Investigaci'on en Computaci'on, Instituto Polit'ecnico Nacional, M'exico City, M'exico.}

\author{U.~Cotti}
\affiliation{Universidad Michoacana de San Nicolás de Hidalgo, Morelia, Mexico }

\author{J.~Cotzomi}
\affiliation{Facultad de Ciencias F'{i}sico Matemáticas, Benemérita Universidad Autónoma de Puebla, Puebla, Mexico }

\author{S.~Coutiño de León}
\affiliation{Department of Physics, University of Wisconsin-Madison, Madison, WI, USA }

\author{E.~De la Fuente}
\affiliation{Departamento de F'{i}sica, Centro Universitario de Ciencias Exactase Ingenierias, Universidad de Guadalajara, Guadalajara, Mexico }

\author{R.~Diaz Hernandez}
\affiliation{Instituto Nacional de Astrof'{i}sica, Óptica y Electrónica, Puebla, Mexico }

\author{B.L.~Dingus}
\affiliation{Physics Division, Los Alamos National Laboratory, Los Alamos, NM 87544, USA }\affiliation{Department of Physics, University of Maryland, College Park, MD, USA }

\author{M.A.~DuVernois}
\affiliation{Department of Physics, University of Wisconsin-Madison, Madison, WI, USA }

\author{M.~Durocher}
\affiliation{Physics Division, Los Alamos National Laboratory, Los Alamos, NM 87544, USA }

\author{J.C.~Díaz-Vélez}
\affiliation{Departamento de F'{i}sica, Centro Universitario de Ciencias Exactase Ingenierias, Universidad de Guadalajara, Guadalajara, Mexico }

\author{R.W.~Ellsworth}
\affiliation{Department of Physics, University of Maryland, College Park, MD, USA }

\author{K.~Engel}
\affiliation{Department of Physics, University of Maryland, College Park, MD, USA }

\author{C.~Espinoza}
\affiliation{Instituto de F'{i}sica, Universidad Nacional Autónoma de México, Ciudad de Mexico, Mexico }

\author{K.L.~Fan}
\affiliation{Department of Physics, University of Maryland, College Park, MD, USA }

\author{K.~Fang}
\affiliation{Department of Physics, University of Wisconsin-Madison, Madison, WI, USA }

\author{M.~Fernández Alonso}
\affiliation{Department of Physics, Pennsylvania State University, University Park, PA, USA }

\author{H.~Fleischhack}
\affiliation{Department of Physics, Catholic University of America, 620 Michigan Avenue NE, Washington, DC 20064}\affiliation{NASA Goddard Space Flight Center, Greenbelt, MD 20771, USA}\affiliation{Center for Research and Exploration in Space Science and Technology, NASA/GSFC, Greenbelt, MD 20771}

\author{N.~Fraija}
\affiliation{Instituto de Astronom'{i}a, Universidad Nacional Autónoma de México, Ciudad de Mexico, Mexico }

\author{J.A.~García-González}
\affiliation{Tecnologico de Monterrey, Escuela de Ingenier\'{i}a y Ciencias, Avenue Eugenio Garza Sada 2501, Monterrey, N.L., Mexico, 64849}

\author{F.~Garfias}
\affiliation{Instituto de Astronom'{i}a, Universidad Nacional Autónoma de México, Ciudad de Mexico, Mexico }

\author{M.M.~González}
\affiliation{Instituto de Astronom'{i}a, Universidad Nacional Autónoma de México, Ciudad de Mexico, Mexico }

\author{J.A.~Goodman}
\affiliation{Department of Physics, University of Maryland, College Park, MD, USA }

\author{J.P.~Harding}
\affiliation{Physics Division, Los Alamos National Laboratory, Los Alamos, NM 87544, USA }

\author{S.~Hernandez}
\affiliation{Instituto de F'{i}sica, Universidad Nacional Autónoma de México, Ciudad de Mexico, Mexico }

\author{J.~Hinton}
\affiliation{Max-Planck Institute for Nuclear Physics, 69117 Heidelberg, Germany}

\author{D.~Huang}
\affiliation{Department of Physics, Michigan Technological University, Houghton, MI, USA }

\author{F.~Hueyotl-Zahuantitla}
\affiliation{Universidad Autónoma de Chiapas, Tuxtla Gutiérrez, Chiapas, México}

\author{P.~Hüntemeyer}
\affiliation{Department of Physics, Michigan Technological University, Houghton, MI, USA }

\author{A.~Iriarte}
\affiliation{Instituto de Astronom'{i}a, Universidad Nacional Autónoma de México, Ciudad de Mexico, Mexico }

\author{V.~Joshi}
\affiliation{Erlangen Centre for Astroparticle Physics, Friedrich-Alexander-Universit\"at Erlangen-N\"urnberg, Erlangen, Germany}

\author{S.~Kaufmann}
\affiliation{Universidad Politecnica de Pachuca, Pachuca, Hgo, Mexico }

\author{J.~Lee}
\affiliation{University of Seoul, Seoul, Rep. of Korea}

\author{J.T.~Linnemann}
\affiliation{Department of Physics and Astronomy, Michigan State University, East Lansing, MI, USA }

\author{A.L.~Longinotti}
\affiliation{Instituto de Astronom'{i}a, Universidad Nacional Autónoma de México, Ciudad de Mexico, Mexico }

\author{G.~Luis-Raya}
\affiliation{Universidad Politecnica de Pachuca, Pachuca, Hgo, Mexico }

\author{K.~Malone}
\affiliation{Space Science and Applications Group, Los Alamos National Laboratory, Los Alamos, NM 87544, USA }

\author{O.~Martinez}
\affiliation{Facultad de Ciencias F'{i}sico Matemáticas, Benemérita Universidad Autónoma de Puebla, Puebla, Mexico }

\author{J.~Martínez-Castro}
\affiliation{Centro de Investigaci'on en Computaci'on, Instituto Polit'ecnico Nacional, M'exico City, M'exico.}

\author{J.A.~Matthews}
\affiliation{Dept of Physics and Astronomy, University of New Mexico, Albuquerque, NM, USA }

\author{P.~Miranda-Romagnoli}
\affiliation{Universidad Autónoma del Estado de Hidalgo, Pachuca, Mexico }

\author{J.A.~Morales-Soto}
\affiliation{Universidad Michoacana de San Nicolás de Hidalgo, Morelia, Mexico }

\author{E.~Moreno}
\affiliation{Facultad de Ciencias F'{i}sico Matemáticas, Benemérita Universidad Autónoma de Puebla, Puebla, Mexico }

\author{M.~Mostafá}
\affiliation{Department of Physics, Pennsylvania State University, University Park, PA, USA }

\author{A.~Nayerhoda}
\affiliation{Instytut Fizyki Jadrowej im Henryka Niewodniczanskiego Polskiej Akademii Nauk, IFJ-PAN, Krakow, Poland }

\author{L.~Nellen}
\affiliation{Instituto de Ciencias Nucleares, Universidad Nacional Autónoma de Mexico, Ciudad de Mexico, Mexico }

\author{M.U.~Nisa \orcid{0000-0002-6859-3944}}\email{Corresponding Author: nisamehr@msu.edu}
\affiliation{Department of Physics and Astronomy, Michigan State University, East Lansing, MI, USA }

\author{R.~Noriega-Papaqui}
\affiliation{Universidad Autónoma del Estado de Hidalgo, Pachuca, Mexico }

\author{L.~Olivera-Nieto}
\affiliation{Max-Planck Institute for Nuclear Physics, 69117 Heidelberg, Germany}

\author{N.~Omodei}
\affiliation{Department of Physics, Stanford University: Stanford, CA 94305–4060, USA}

\author{Y.~Pérez Araujo}
\affiliation{Instituto de Astronom'{i}a, Universidad Nacional Autónoma de México, Ciudad de Mexico, Mexico }

\author{E.G.~Pérez-Pérez}
\affiliation{Universidad Politecnica de Pachuca, Pachuca, Hgo, Mexico }

\author{C.D.~Rho}
\affiliation{Department of Physics, Sungkyunkwan University, Suwon 16419, South Korea}

\author{D.~Rosa-González \orcid{0000-0003-1327-0838}}
\affiliation{Instituto Nacional de Astrof'{i}sica, Óptica y Electrónica, Puebla, Mexico }

\author{E.~Ruiz-Velasco}
\affiliation{Max-Planck Institute for Nuclear Physics, 69117 Heidelberg, Germany}

\author{H.~Salazar}
\affiliation{Facultad de Ciencias F'{i}sico Matemáticas, Benemérita Universidad Autónoma de Puebla, Puebla, Mexico }

\author{D.~Salazar-Gallegos}
\affiliation{Department of Physics and Astronomy, Michigan State University, East Lansing, MI, USA }

\author{A.~Sandoval}
\affiliation{Instituto de F'{i}sica, Universidad Nacional Autónoma de México, Ciudad de Mexico, Mexico }

\author{M.~Schneider}
\affiliation{Department of Physics, University of Maryland, College Park, MD, USA }

\author{J.~Serna-Franco}
\affiliation{Instituto de F'{i}sica, Universidad Nacional Autónoma de México, Ciudad de Mexico, Mexico }

\author{A.J.~Smith}
\affiliation{Department of Physics, University of Maryland, College Park, MD, USA }

\author{Y.~Son}
\affiliation{University of Seoul, Seoul, Rep. of Korea}

\author{R.W.~Springer}
\affiliation{Department of Physics and Astronomy, University of Utah, Salt Lake City, UT, USA }

\author{O.~Tibolla}
\affiliation{Universidad Politecnica de Pachuca, Pachuca, Hgo, Mexico }

\author{K.~Tollefson}
\affiliation{Department of Physics and Astronomy, Michigan State University, East Lansing, MI, USA }

\author{I.~Torres}
\affiliation{Instituto Nacional de Astrof'{i}sica, Óptica y Electrónica, Puebla, Mexico }

\author{R.~Torres-Escobedo}
\affiliation{Tsung-Dao Lee Institute \& School of Physics and Astronomy, Shanghai Jiao Tong University, Shanghai, People's Republic of China}

\author{R.~Turner}
\affiliation{Department of Physics, Michigan Technological University, Houghton, MI, USA }

\author{F.~Ureña-Mena}
\affiliation{Instituto Nacional de Astrof'{i}sica, Óptica y Electrónica, Puebla, Mexico }

\author{E.~Varela}
\affiliation{Facultad de Ciencias F'{i}sico Matemáticas, Benemérita Universidad Autónoma de Puebla, Puebla, Mexico }

\author{L.~Villaseñor}
\affiliation{Facultad de Ciencias F'{i}sico Matemáticas, Benemérita Universidad Autónoma de Puebla, Puebla, Mexico }

\author{X.~Wang}
\affiliation{Department of Physics, Michigan Technological University, Houghton, MI, USA }

\author{I.J.~Watson}
\affiliation{University of Seoul, Seoul, Rep. of Korea}

\author{E.~Willox}
\affiliation{Department of Physics, University of Maryland, College Park, MD, USA }

\author{S.~Yun-Cárcamo}
\affiliation{Department of Physics, University of Maryland, College Park, MD, USA }

\author{H.~Zhou}
\affiliation{Tsung-Dao Lee Institute \& School of Physics and Astronomy, Shanghai Jiao Tong University, Shanghai, People's Republic of China}

\author{C.~de León}
\affiliation{Universidad Michoacana de San Nicolás de Hidalgo, Morelia, Mexico }

\collaboration{HAWC Collaboration}
\author{J. F. Beacom}
%\email{beacom.7@osu.edu}
%\thanks{\scriptsize \!\!  \href{http://orcid.org/0000-0002-0005-2631}{orcid.org/0000-0002-0005-2631}}
\affiliation{Center for Cosmology and AstroParticle Physics (CCAPP), Ohio State University, Columbus, Ohio 43210, USA}
\affiliation{Department of Physics, Ohio State University, Columbus, Ohio 43210, USA}
\affiliation{Department of Astronomy, Ohio State University, Columbus, Ohio 43210, USA}

\author{T. Linden}
%\email{linden.70@osu.edu}
%\thanks{\scriptsize \!\!  \href{http://orcid.org/0000-0001-9888-0971}{orcid.org/0000-0001-9888-0971}}
\affiliation{The Oskar Klein Centre, Department of Physics, Stockholm University, AlbaNova, SE-10691 Stockholm, Sweden}

\author{K. C. Y. Ng}
%\email{ chun-yu.ng@weizmann.ac.il}
%\thanks{\scriptsize \!\! \href{http://orcid.org/0000-0001-8016-2170}{orcid.org/0000-0001-8016-2170}}
\affiliation{Department of Physics, The Chinese University of Hong Kong, Sha Tin, Hong Kong, China}

\author{A. H. G. Peter}
%\email{apeter@physics.osu.edu}
%\thanks{ \scriptsize \!\!  \href{http://orcid.org/0000-0002-8040-6785}{orcid.org/0000-0002-8040-6785}}
\affiliation{Center for Cosmology and AstroParticle Physics (CCAPP), Ohio State University, Columbus, Ohio 43210, USA}
\affiliation{Department of Physics, Ohio State University, Columbus, Ohio 43210, USA}
\affiliation{Department of Astronomy, Ohio State University, Columbus, Ohio 43210, USA}
\affiliation{School of Natural Sciences, Institute for Advanced Study, 1 Einstein Drive, Princeton, NJ 08540}

\author{B. Zhou \vspace{0.5cm}}
%\email{zhou.1877@osu.edu}
%\thanks{\scriptsize \!\!  \href{http://orcid.org/0000-0003-1600-8835}{orcid.org/0000-0003-1600-8835}}
\affiliation{William H. Miller III Department of Physics and Astronomy, Johns Hopkins University, Baltimore, Maryland 21218, USA}

\hfill
\date{\today}

\begin{abstract}

We report the first detection of a TeV gamma-ray flux from the solar disk (6.3$\sigma$), based on 6.1 years of data from the High Altitude Water Cherenkov (HAWC) observatory.  The 0.5--2.6 TeV spectrum is well fit by a power law, dN/dE = $A (E/1 \text{ TeV})^{-\gamma}$, with $A = (1.6 \pm 0.3) \times 10^{-12}$ TeV$^{-1}$ cm$^{-2}$ s$^{-1}$ and $\gamma = -3.62 \pm 0.14$.  The flux shows a strong indication of anticorrelation with solar activity.  These results extend the bright, hard GeV emission from the disk observed with Fermi-LAT, seemingly due to hadronic Galactic cosmic rays showering on nuclei in the solar atmosphere. However, current theoretical models are unable to explain the details of how solar magnetic fields shape these interactions. HAWC's TeV detection thus deepens the mysteries of the solar-disk emission.

\end{abstract}

\maketitle

%%%%%%%%%%%%%%%%%%%%%%%%%%%%%Section1%%%%%%%%%%%%%%%%%%%%%%%%%%%%%%%%%%%
%%%%%%%%%%%%%%%%%%%%%%%%%%%%%%%%%%%%%%%%%%%%%%%%%%%%%%%%%%%%%%%%%%%%%%%%
%TC:endignore

The Sun is one of the most widely studied sources in multi-messenger astrophysics. It can be probed in detail through direct observations across the electromagnetic spectrum, in MeV-scale neutrinos, and in accelerated particles, as well as indirectly through helioseismology, its cosmic-ray shadow, and magnetic field measurements~\cite{HelioDecadal, AstroDecadal}.

However, the Sun's emission at high energies remains mysterious.  For example, the solar disk is a bright, continuous source of gamma rays, with Fermi-LAT observations~\cite{2011ApJ734116A, Ng:2015gya, Linden:2018exo, Tang:2018wqp, Linden:2020lvz} (building on earlier hints from EGRET~\cite{2008A&A...480..847O}) showing gamma-ray emission between 0.1--200 GeV and revealing several puzzling features.  The primary emission mechanism seems to be the decay of $\pi^0$ produced by the scattering of hadronic Galactic cosmic rays with nuclei in the solar atmosphere over the full disk, with the requirement that the cosmic rays must first be converted from incoming to outgoing by magnetic fields~\cite{Seckel:1991ffa}. Without this magnetic redirection, cosmic rays would only be grazing the surface of the Sun, encountering limited column density and the disk emission would be much fainter ~\cite{Zhou:2016ljf}. Even so, compared to theoretical expectations~\cite{Seckel:1991ffa, Zhou:2016ljf, Li:2020gch, Mazziotta:2020uey, Gutierrez:2019fna, Gutierrez:2022mor}, the observed flux of gamma rays from the solar disk is brighter and the spectrum is harder (with an unexplained dip near 40 GeV) \cite{2011ApJ734116A,  Ng:2015gya, Tang:2018wqp}.  Additionally, the flux is anticorrelated with solar activity and the emission across the disk is nonuniform~\cite{Linden:2018exo,  Linden:2020lvz}.

Decisive new probes are needed to solve these puzzles, and observations at high energies are especially important, for multiple reasons.  Unlike the cosmic-ray spectrum, which falls as $\sim E^{-2.7}$, the solar-minimum gamma-ray spectrum (except for the dip) falls as $\sim E^{-2.2}$ up to at least 200 GeV. This trend must eventually reach a break energy, where cosmic rays are no longer sufficiently deflected in the Sun's magnetic field, and will be an important clue to the details of their propagation.  Separately, the highest-energy gamma rays are likely produced at the greatest depths under the photosphere ($\sim 1000$ km), thus providing sensitivity to otherwise-hidden magnetic fields.  Last, understanding the solar emission will be important for tests of new physics, including dark matter~\cite{Arguelles:2017eao, Ng:2017aur, Edsjo:2017kjk, Leane:2017vag, HAWC:2018szf}.

In this {\it Letter}, we use observations with the High Altitude Water Cherenkov (HAWC) observatory to probe the solar disk in the TeV range.  We substantially improve upon the earlier HAWC search that set an upper limit on the gamma-ray flux~\cite{HAWC:2018rpf} (ARGO-YBJ also set a limit~\cite{ARGO-YBJ:2019mdq}).  As described below, here we use a larger dataset, better reconstruction algorithms, and new signal-isolation techniques.  In the following, we describe the HAWC data, our analysis methods, tests of the time variation and spectrum slope, and then conclude.  Further details are given in Supplemental Material (S.M.).

%%%%%%%%%%%%%%%%%%%%%%%%%%%%%Section2%%%%%%%%%%%%%%%%%%%%%%%%%%%%%%%%%%%
%%%%%%%%%%%%%%%%%%%%%%%%%%%%%%%%%%%%%%%%%%%%%%%%%%%%%%%%%%%%%%%%%%%%%%%%

\emph{\textbf{HAWC Data.---}}
The HAWC observatory, located at an altitude of 4100 m near Puebla, Mexico, is designed to detect multi-TeV cosmic rays and gamma rays through atmospheric-shower secondary particles that reach the ground~\cite{2017ApJ...843...39A,  HAWC:2020hrt}.  These secondaries are detected in an array of 300 detector units that employ the water-Cherenkov technique and operate near-continuously.  The vast majority of detected showers are induced by hadronic cosmic rays, which cause a near-isotropic and near-constant background.  In searches for gamma-ray-induced showers, the cosmic-ray background can be greatly reduced (to a fraction $10^{-1}$ to $10^{-4}$, depending on the energy of the primary particle) and any gamma-ray signals nearly perfectly preserved, by cuts based on shower topology (gamma-ray showers are compact, while hadronic showers have a broader and clumpier footprint)~\cite{2017ApJ...843...39A}.  The angular resolution of HAWC depends on energy and zenith angle, ranging from $\sim 1^\circ$ at 1 TeV to $0.2^\circ$ above 10 TeV (see S.M.).  HAWC's gamma-ray observations, which cover the entire sky at zenith angles $0$--$45^\circ$, offer excellent sensitivity to both source and diffuse emission, as has been exploited in a variety of studies (see, e.g., Refs.~\cite{HAWC:2020hrt, 2017ApJ...842...85A, HAWC:2019tcx,2021ApJ...917....6A, Albert:2021cwz,HAWC:2018wju, 2017arXiv171000890H}).

HAWC is among the few detectors capable of observing the Sun in the TeV range~\cite{HAWC:2018rpf}.  Its large field of view and high livetime fraction allow continuous exposure as the Sun is tracked across the sky. Compared to the earlier HAWC analysis, here we make significant improvements.  First, we use a larger and more varied exposure, spanning November 2014 to January 2021 (6.1 years). The first half of the data corresponds to an active but declining part of solar cycle 24, while the second half corresponds to the minimum of solar cycle 25; this long baseline thus allows for tests of time variability. Second, we use an improved offline reconstruction sample (Pass 5, compared to Pass 4 in previous work \cite{HAWC:2018rpf}) with new calibrations and better algorithms.  The new data have superior angular resolution and background rejection, particularly at low energies, improving the sensitivity by a factor of $\sim$2--5, depending on the source spectrum (see Fig.~\ref{fig:sensi} in S.M.).  Third, we use a new, data-driven approach to separate the gamma-ray signal from backgrounds, taking into account the suppression of cosmic-ray fluxes from directions near the Sun (the shadow effect).

%%%%%%%%%%%%%%%%%%%%%%%%%%%%%Section3%%%%%%%%%%%%%%%%%%%%%%%%%%%%%%%%%%%
%%%%%%%%%%%%%%%%%%%%%%%%%%%%%%%%%%%%%%%%%%%%%%%%%%%%%%%%%%%%%%%%%%%%%%%%

%%%%%%%%%%%%%%%%%%%%%%%%%%%%FIGURE1%%%%%%%%%%%%%%%%%%%%%%%%%%%%%%%%%
\begin{figure*}[t]
\makebox[0.8\width][c]{
\begin{tabular}{@{}cc@{}}
\includegraphics[width=0.47\textwidth]{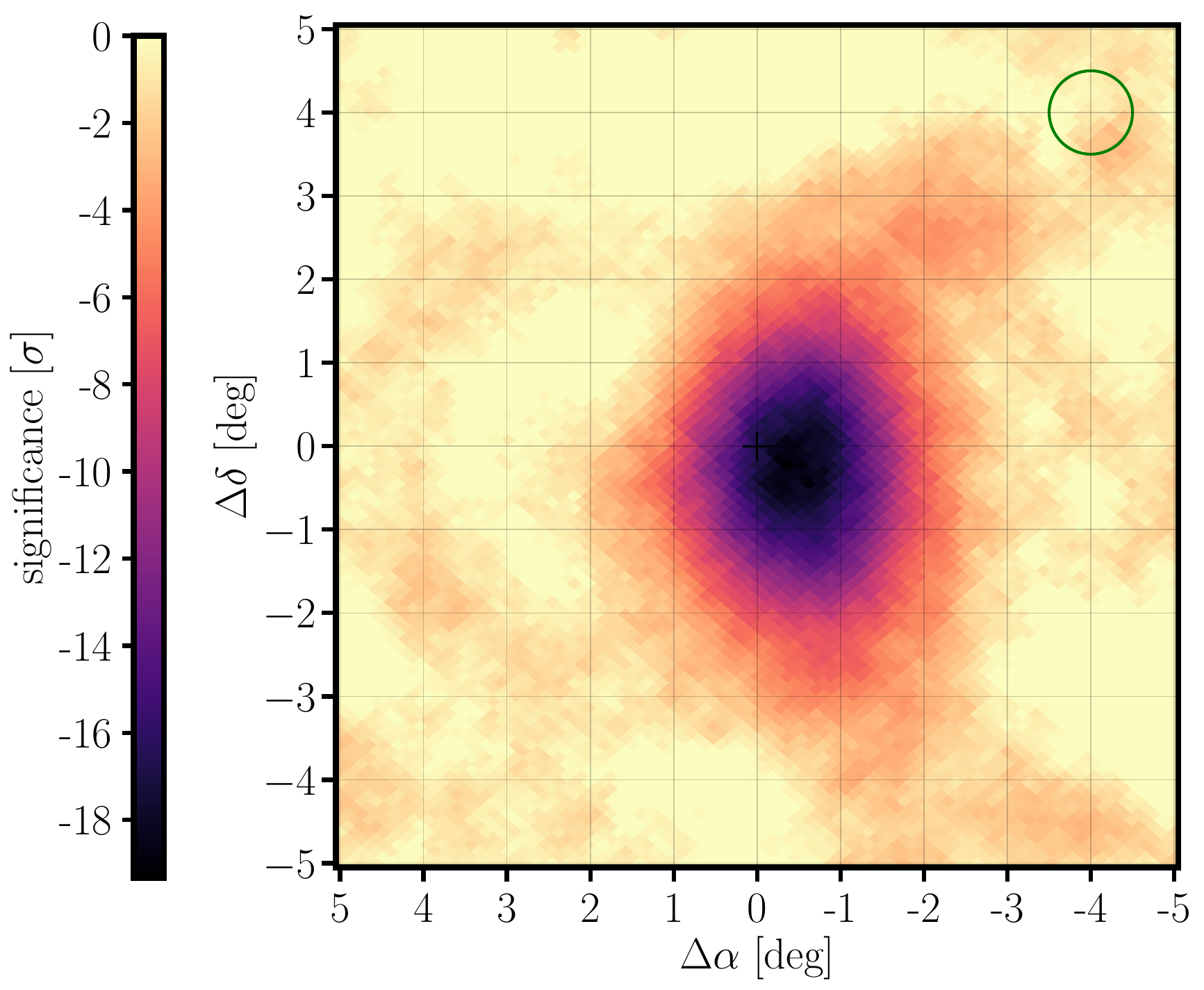} &
\includegraphics[width=0.47\textwidth]{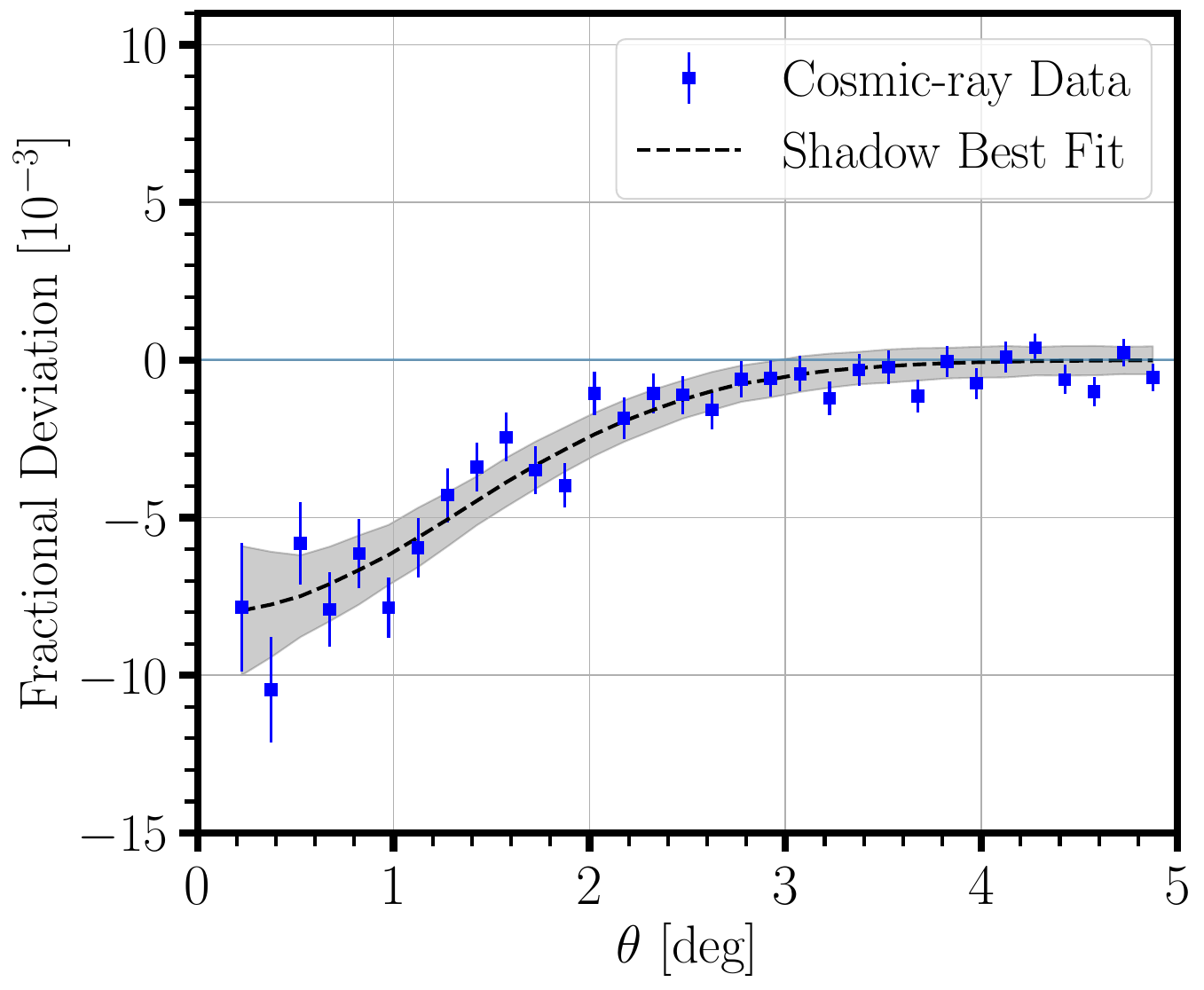} \\
\includegraphics[width=0.47\textwidth]{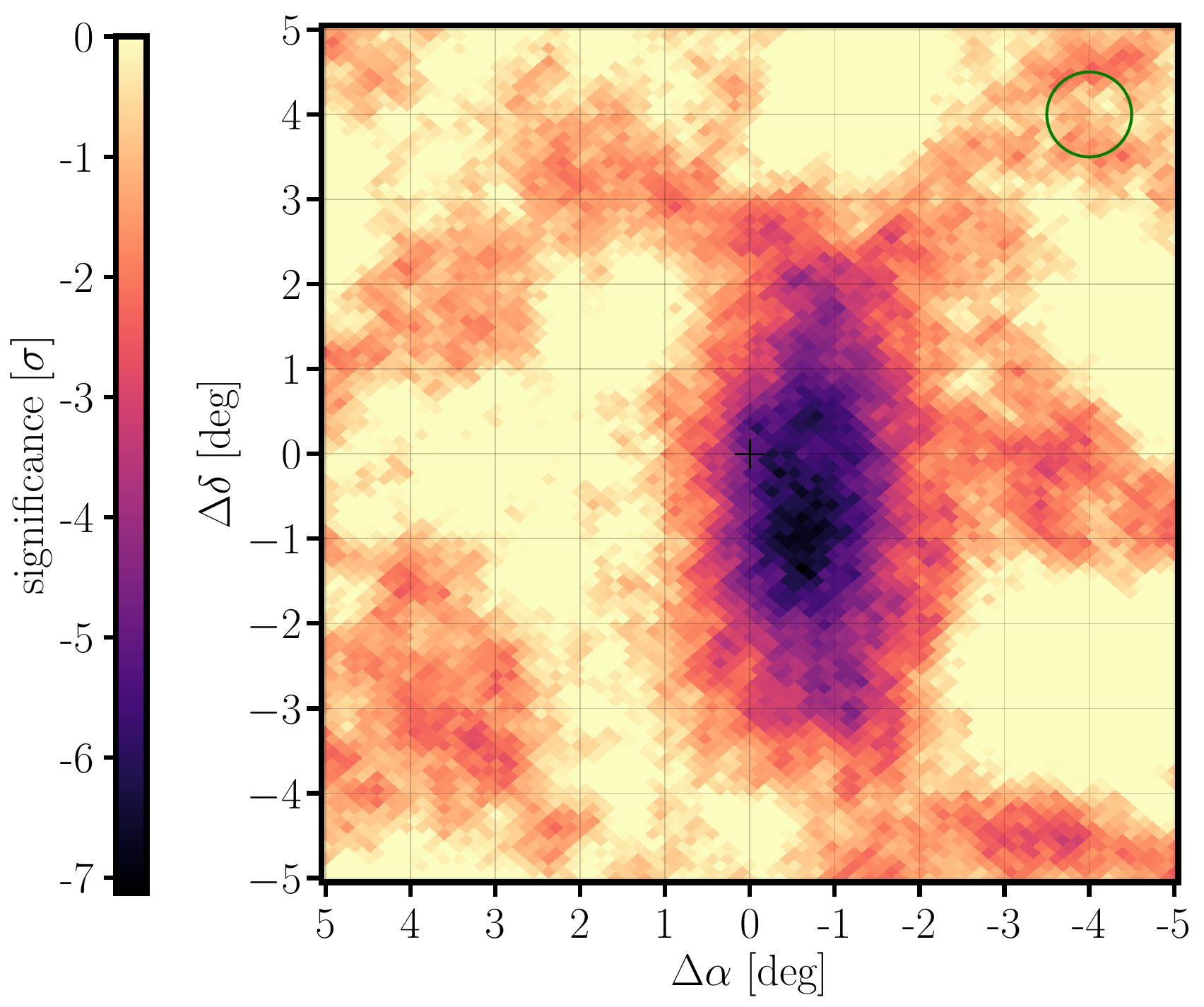} &
 \includegraphics[width=0.47\textwidth]{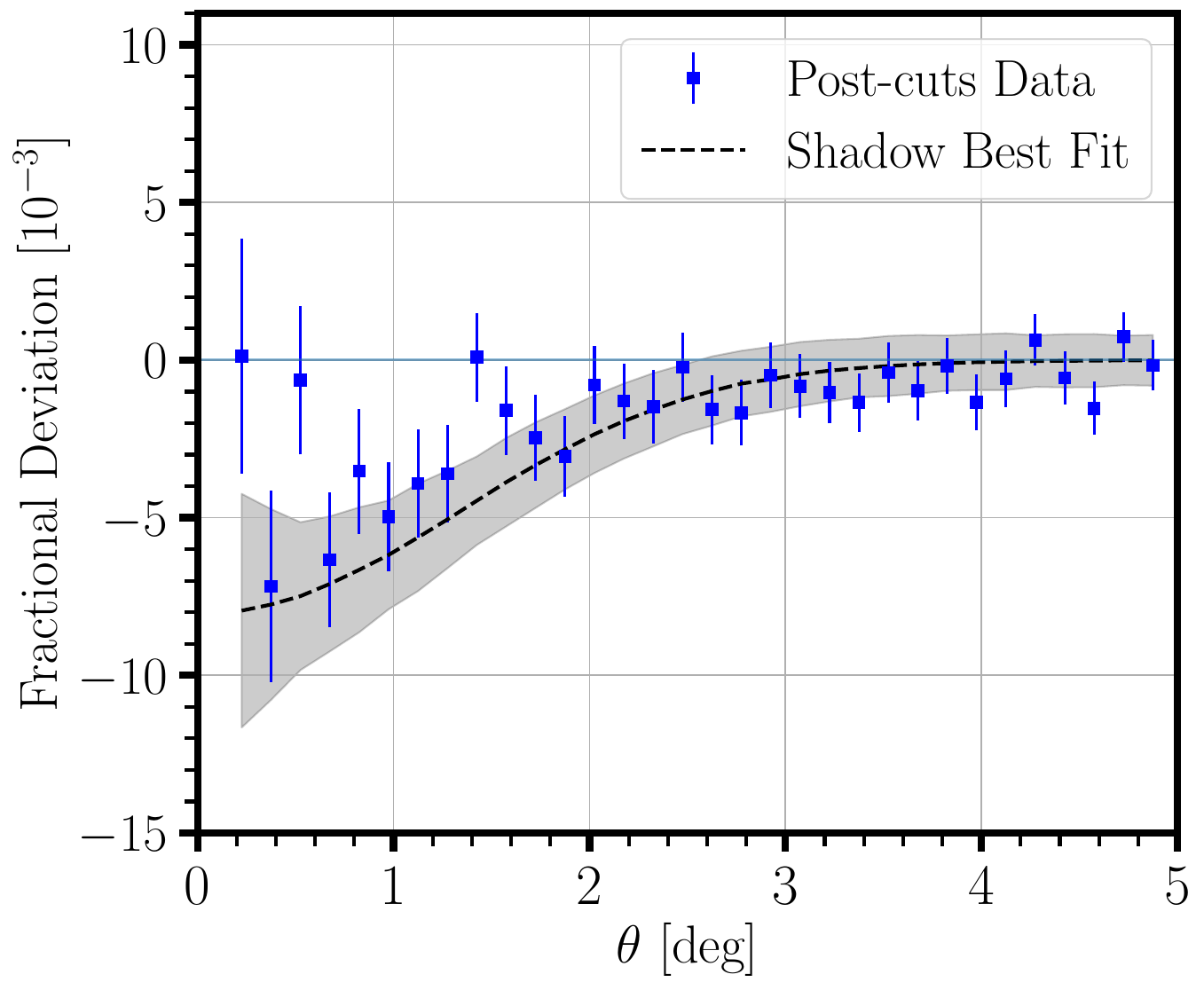}\\
\includegraphics[width=0.47\textwidth]{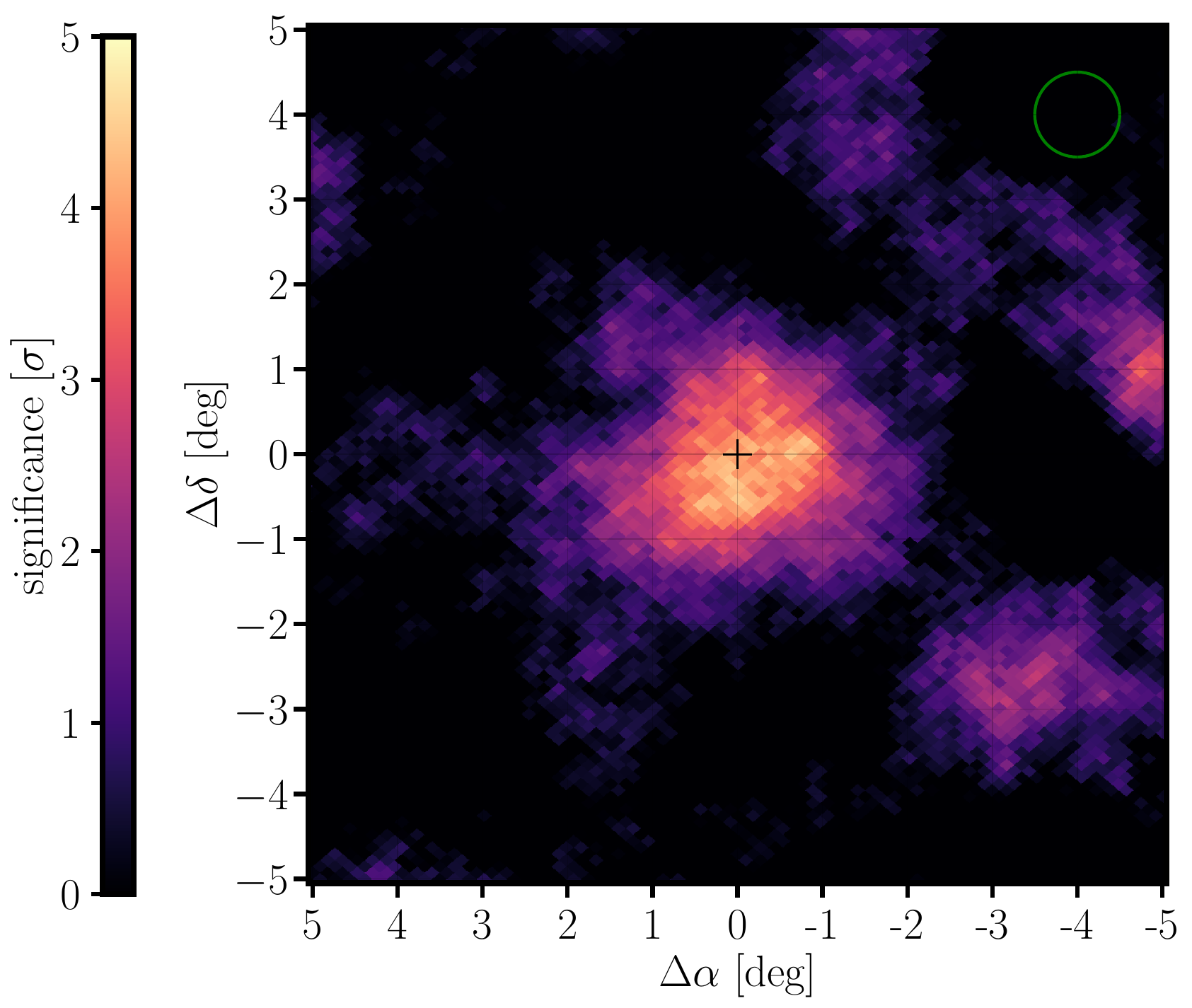} &
\includegraphics[width=0.47\textwidth]{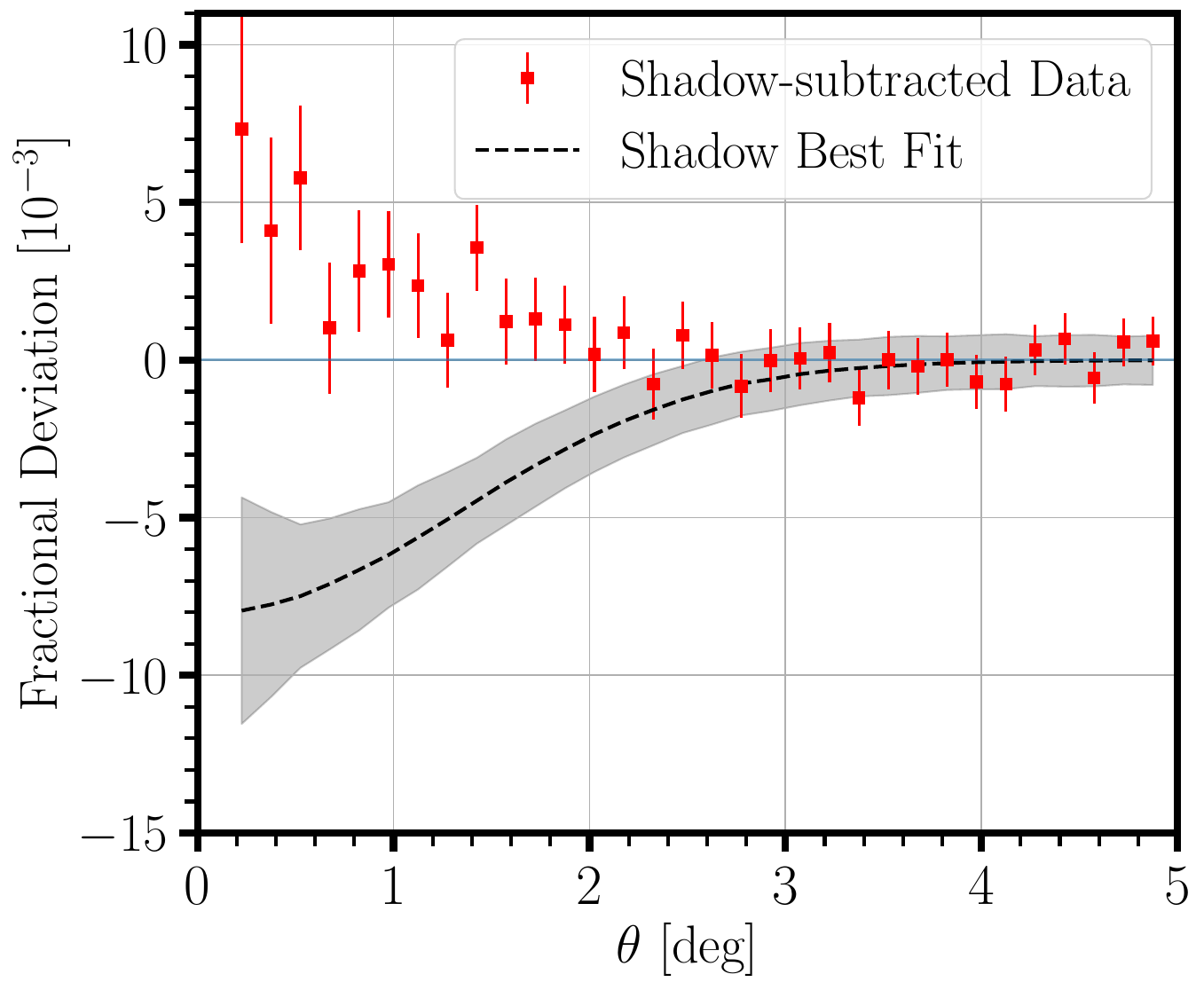}
\end{tabular}
}
\caption{Results for the example of bin B3.  \textbf{Left:} Significance maps in Sun-centered coordinates for 6.1 years of data, smoothed with a 1$^\circ$ top-hat function for visual clarity.  The green circle illustrates the true point spread function. \textbf{Right:} Angular profiles (steps of $0.15^\circ$ from the Sun) of the fractional deviation from background.  The black dashed line shows the projection of the best-fit 2D Gaussian model fitted to the shadow, with the shaded band indicating the total uncertainty in the model.  The \textit{top row} shows the cosmic-ray dominated data. The \textit{middle row} shows the events that survive the gamma-hadron separation cuts. The \textit{bottom row} is after subtracting the measured cosmic-ray shadow (see top-row data) from the middle-row data, leaving a gamma-ray excess at the position of the Sun (marked by a cross).}
\label{fig:maps}
\end{figure*}
%%%%%%%%%%%%%%%%%%%%%%%%%%%%END FIGURE%%%%%%%%%%%%%%%%%%%%%%%%%%%%%%

\emph{\textbf{Main Analysis.---}}
To search for TeV-range gamma-ray emission from the solar disk, we develop a simple but powerful analysis technique that measures the signal and background independently.

To begin, we select only well-reconstructed events with shower cores within the fiducial area of HAWC~\cite{2017arXiv171000890H}.  We cut Milky Way sources and diffuse gamma-ray emission by excluding times when the Sun is within $\pm 10^\circ$ of the Galactic plane.  Extragalactic gamma-ray emission is minor and is smeared into the background as we track the Sun across the sky.  The fluxes of isotropic electron cosmic rays and the directional gamma rays they produce by inverse-Compton scattering of solar photons are both negligible in the TeV range~\cite{Zhou:2016ljf, Moskalenko:2006ta, Orlando:2020ezh}.

We bin the data into 11 analysis bins based on the fraction of the detector array triggered.  Higher-energy events trigger a larger fraction of the array, corresponding to higher-numbered bins; however, the energy resolution ($\simeq 50\%$--$100\%$) is large enough that the bins are partially correlated.  We use Bx to denote analysis bin x.  Table~\ref{tab:bins} in S.M.\ gives the details of these bins.

Following the steps detailed below, we estimate the excess of gamma-ray events at the moving solar position. We obtain the spatial distribution of the data using the background-estimation and skymap-making procedures of Refs.~\cite{2003ApJ...595..803A, HAWC:2018rpf}. We record the data and background counts in equal-area pixels with a mean spacing of $0.11^\circ$ on a HEALpix grid \cite{Gorski:2004by}. Prior to any gamma-hadron separation cuts, the data are dominated by hadronic cosmic rays. In general, a gamma-ray source appears as an excess of events in a particular direction after gamma-hadron separation cuts. Near the Sun, the analysis is complicated by an anisotropy in the background called the ``Sun shadow,''  where some Galactic cosmic-ray trajectories are blocked by the Sun.  The shadow is not perfectly round, and it has a slightly displaced position; these effects are due to deflections of cosmic rays in the Sun's coronal fields and in the interplanetary magnetic field~\cite{2018PhRvL.120c1101A, IceCube:2020bsn}.  Our new analysis --- validated on simulations and on observations of the Moon shadow --- takes the Sun shadow into account, which was a limiting systematic of our previous work \cite{HAWC:2018rpf}.

To isolate any disk signal we must accurately subtract the expected shadow effect, which takes three steps, as we illustrate for the example of B3.  At each step, for every pixel $i$ in the map, we record the number of events $N_{i}$ and report the fractional deviation relative to the background, $\left< N_{i} \right>$, calculating the significance following Ref.~\cite{1983ApJ...272..317L}. The fractional deviation  is  given by $N_i/\left< N \right>_i-1$ and shows the amplitude of the deficit (or excess) in the pixel $i$.
\begin{enumerate}

\item 
Figure~\ref{fig:maps} (top row) shows the map for the cosmic-ray dominated data (before gamma-hadron cuts).  This step measures the shadow's spatial profile and amplitude with high statistics.  Before the gamma-hadron cuts, there are $4.0 \times 10^4$ fewer events within in a $1.1^\circ$ region of interest around the Sun (comparable to the 1-$\sigma$ width of the shadow) than the $5.5 \times 10^6$ expected from the isotropic background.  Figure~\ref{fig:maps} (top row) also shows the angular profile centered on the Sun (not its shadow).     

\item
Figure~\ref{fig:maps} (middle row) shows the same results after gamma-hadron cuts. If there were no shadow, this step would yield our results for the disk emission.  However, the shadow persists because the data is still cosmic-ray dominated, though the shadow significance is less and there may be a positive contribution from disk emission.  There are now $6.7 \times 10^3$ fewer events within the region of interest than the $1.7 \times 10^6$ expected from background. The true shadow profile measured in step one needs to be subtracted from this data to reveal any positive contribution from gamma rays.

\item
Figure~\ref{fig:maps} (bottom row) shows what remains after we subtract (in two dimensions) the re-scaled shadow map measured in the top row from the gamma-hadron cut data in the middle row. The re-scaling takes into account the reduced number of events following the cuts. There is now an excess of $6.3\times 10^3$ events in the region of interest relative to the background. In the absence of a gamma-ray signal, this step should result in a residual consistent with the isotropic background and statistical fluctuations. In the presence of a gamma-ray signal centered on the Sun, there should be an excess relative to the background, with a smoothly falling radial profile. The event counts in each pixel are subject to Poisson errors and are propagated as such during the shadow subtraction.

\end{enumerate}

We repeat these steps for all analysis bins used in this work.  In B2 (median energy 0.6 TeV), we detect a gamma-ray excess at a significance of $4.2\sigma$.  In B3 (1 TeV), the excess is $4.5 \sigma$, and in B4 (1.7 TeV), it is $5.1\sigma$.  These significance values, based on the number of excess gamma-ray events above the isotropic background (not the shadow), are calculated using the Li \& Ma method~\cite{1983ApJ...272..317L}.  The combined significance of the excess in these three bins is 6.3$\sigma$, exceeding requirements for a discovery.  No significant excesses are observed in the lower-energy bins (where the gamma-hadron separation and angular resolution worsen) or the higher-energy bins (where the statistics worsen); further details are given in S.M.

A key advantage of our new analysis technique is that it allows separate measurements of the background (before gamma-hadron cuts) and a potential signal (after gamma-hadron cuts) for the same exposure in terms of sky directions and durations.  Another is that it directly and model-independently measures the shadow from data, without needing any time-dependent theoretical modeling of its complex details.      

A potential systematic effect in our analysis could be over-subtracting the shadow, which would result in an artificial signal.  We perform several cross-checks to test for this possibility, finding no problems.  Further details are given in S.M.

\begin{itemize}

\item For off-Sun regions, we simulate the effects of a shadow, which we subtract following the procedures above.  We find no evidence of spurious gamma-ray sources due to over-subtraction.

\item We repeat this, but now also simulate the effects of a point source of flux $2 \times 10^{-12}$ TeV$^{-1}$ cm$^{-2}$ s$^{-1}$ at 1 TeV and spectrum falling as $E^{-3}$, placed within the simulated shadow.  We find that we can recover this source with significance $> 6\sigma$.
    
\item HAWC observes a significant cosmic-ray shadow for the Moon~\cite{Abeysekara:2018syp}. We repeat the entire analysis using data around the Moon and find no evidence of gamma-ray emission.
    
\end{itemize}

%%%%%%%%%%%%%%%%%%%%%%%%%%%%%Section4%%%%%%%%%%%%%%%%%%%%%%%%%%%%%%%%%%%
%%%%%%%%%%%%%%%%%%%%%%%%%%%%%%%%%%%%%%%%%%%%%%%%%%%%%%%%%%%%%%%%%%%%%%%%

\emph{\textbf{Time Variation.---}}
We test for time dependence in the signal by analyzing the data split into two halves: Nov.\ 2014 to Dec.\ 2017 (closer to solar maximum) and Jan.\ 2018 to Jan.\ 2021 (nearly matching solar minimum).

\begin{figure*}[b!]
\makebox[1.00\width][c]{
\begin{tabular}{@{}ccc@{}}
\includegraphics[width=0.30\textwidth]{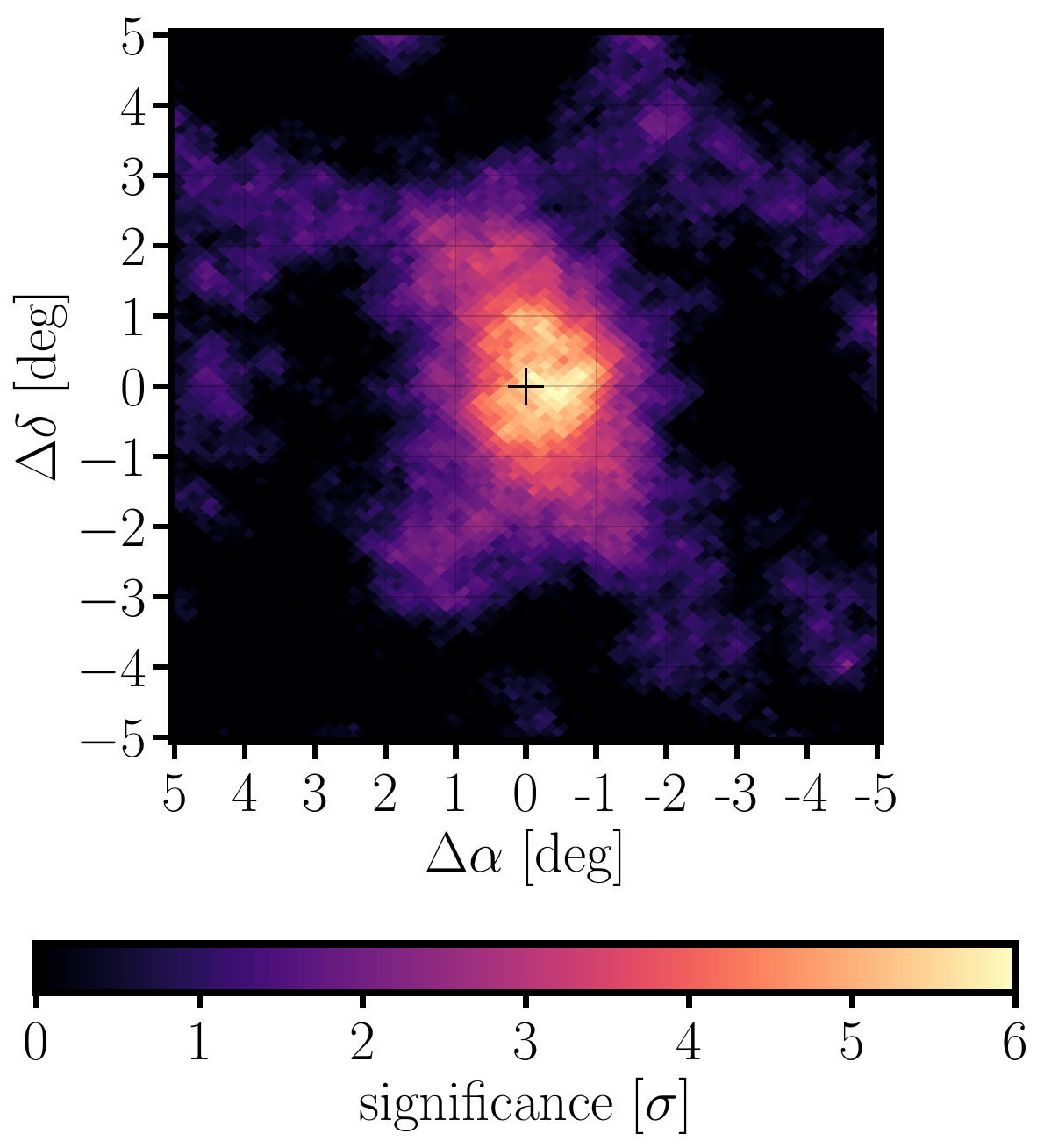}&
\includegraphics[width=0.30\textwidth]{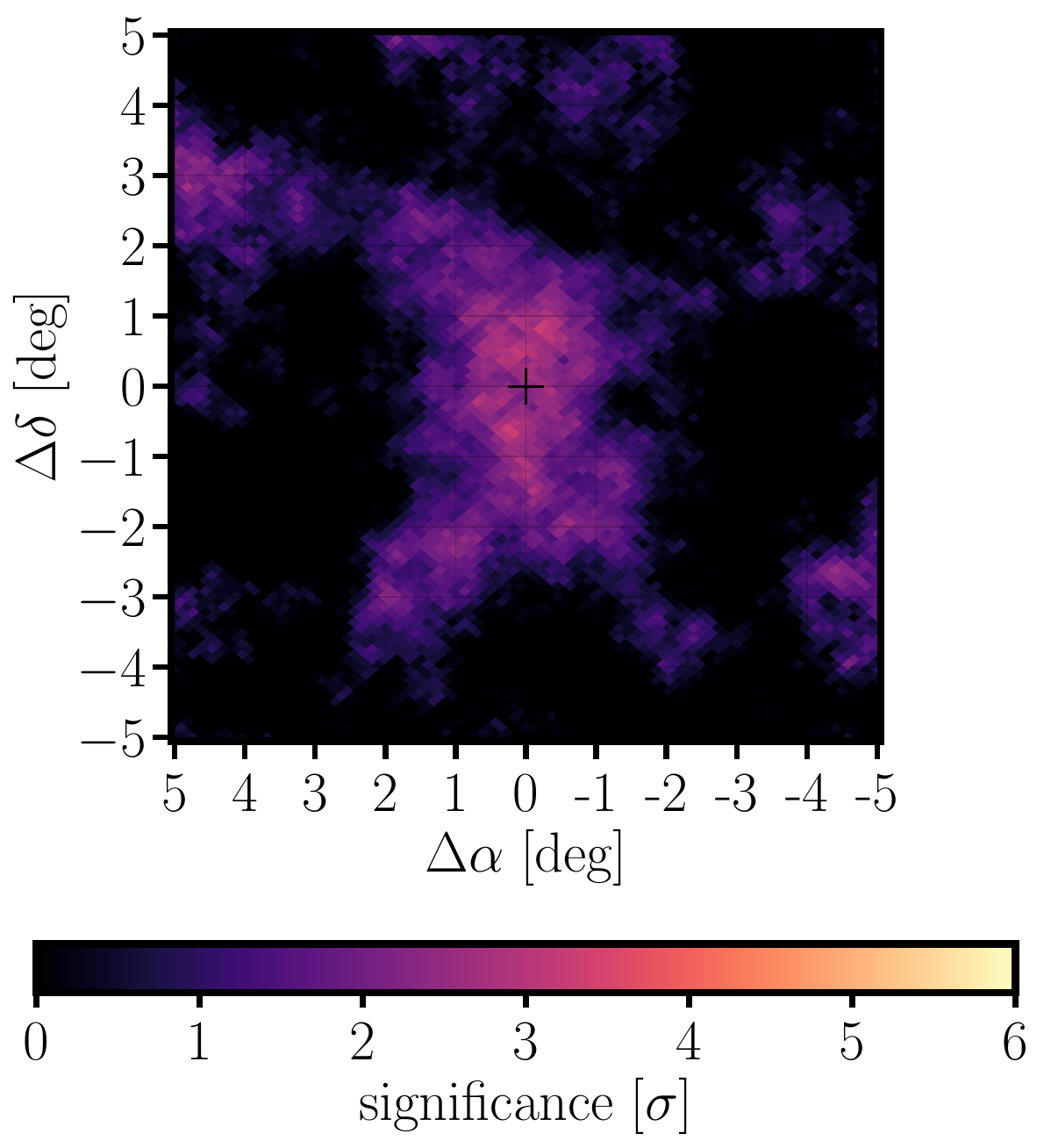} &
\includegraphics[width=0.30\textwidth]{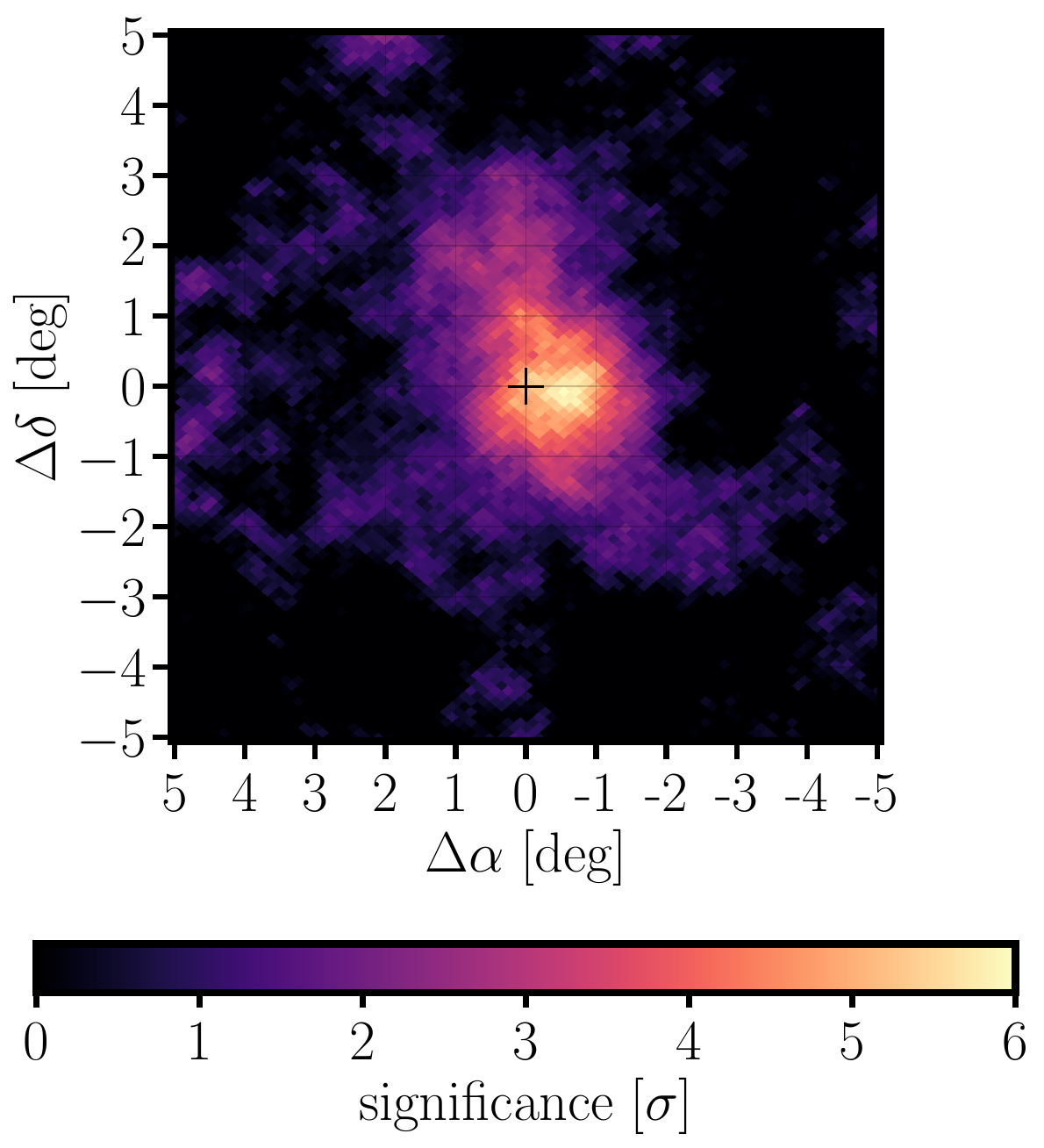}\\ 
\includegraphics[width=0.30\textwidth]{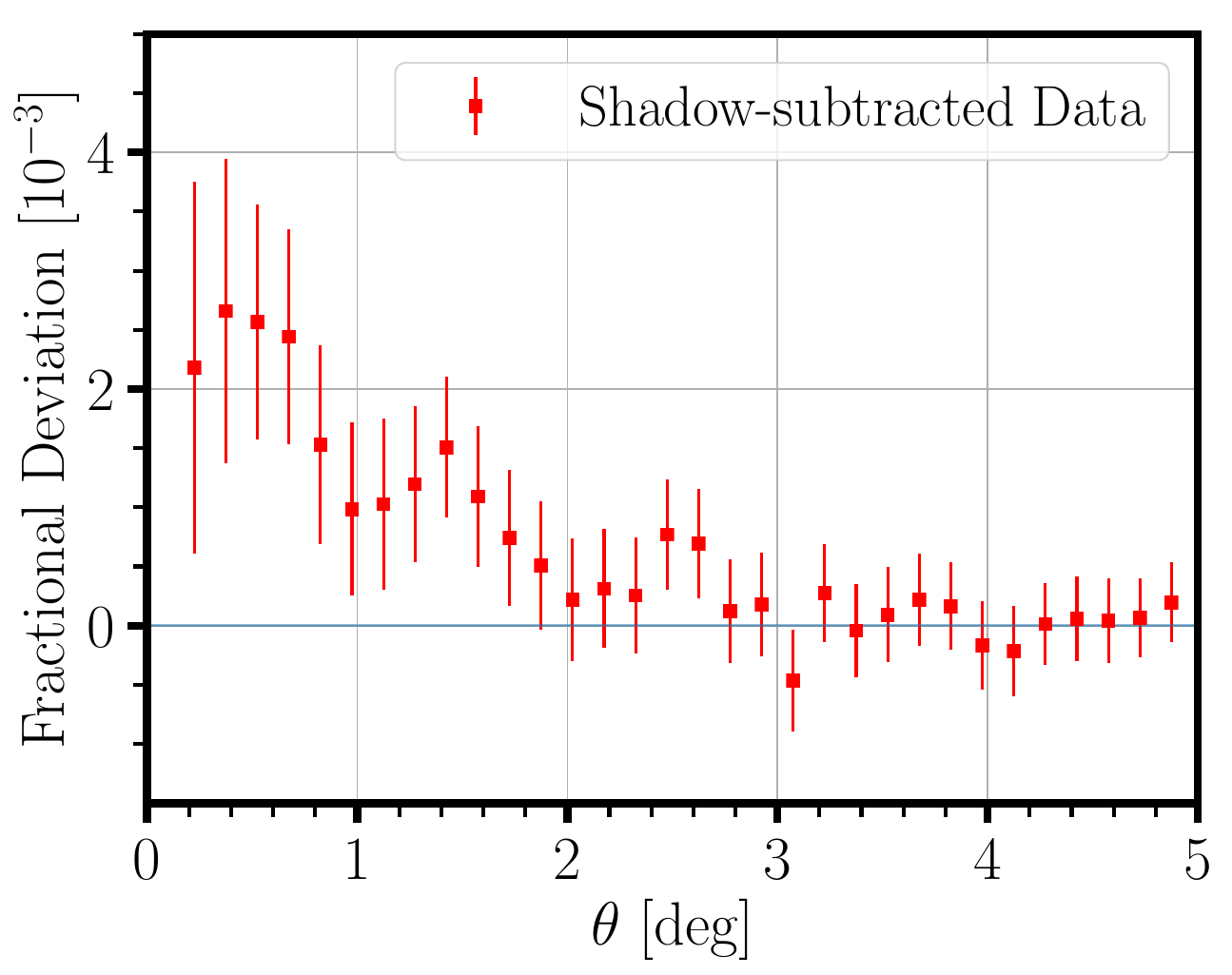} &
\includegraphics[width=0.30\textwidth]{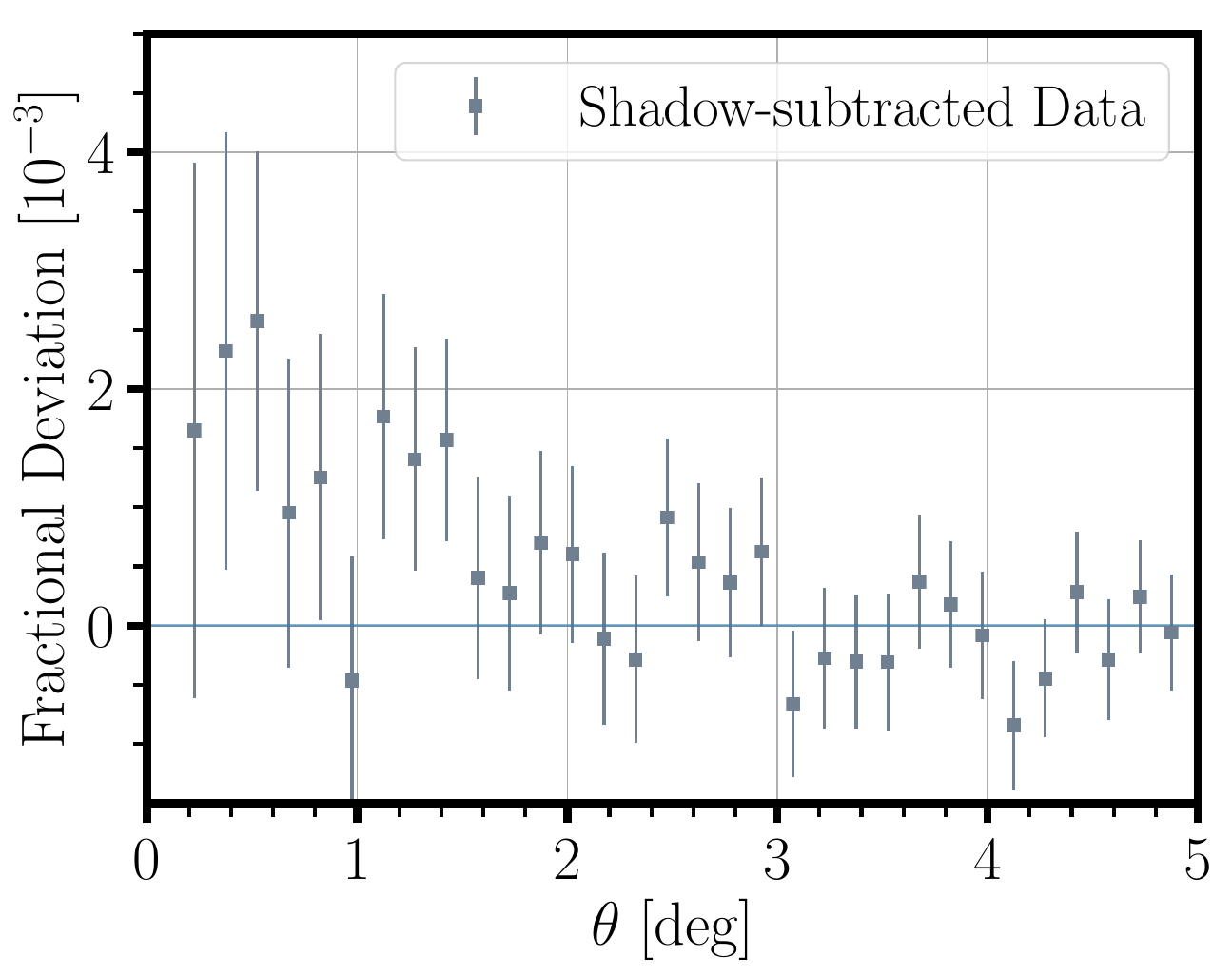} &
\includegraphics[width=0.30\textwidth]{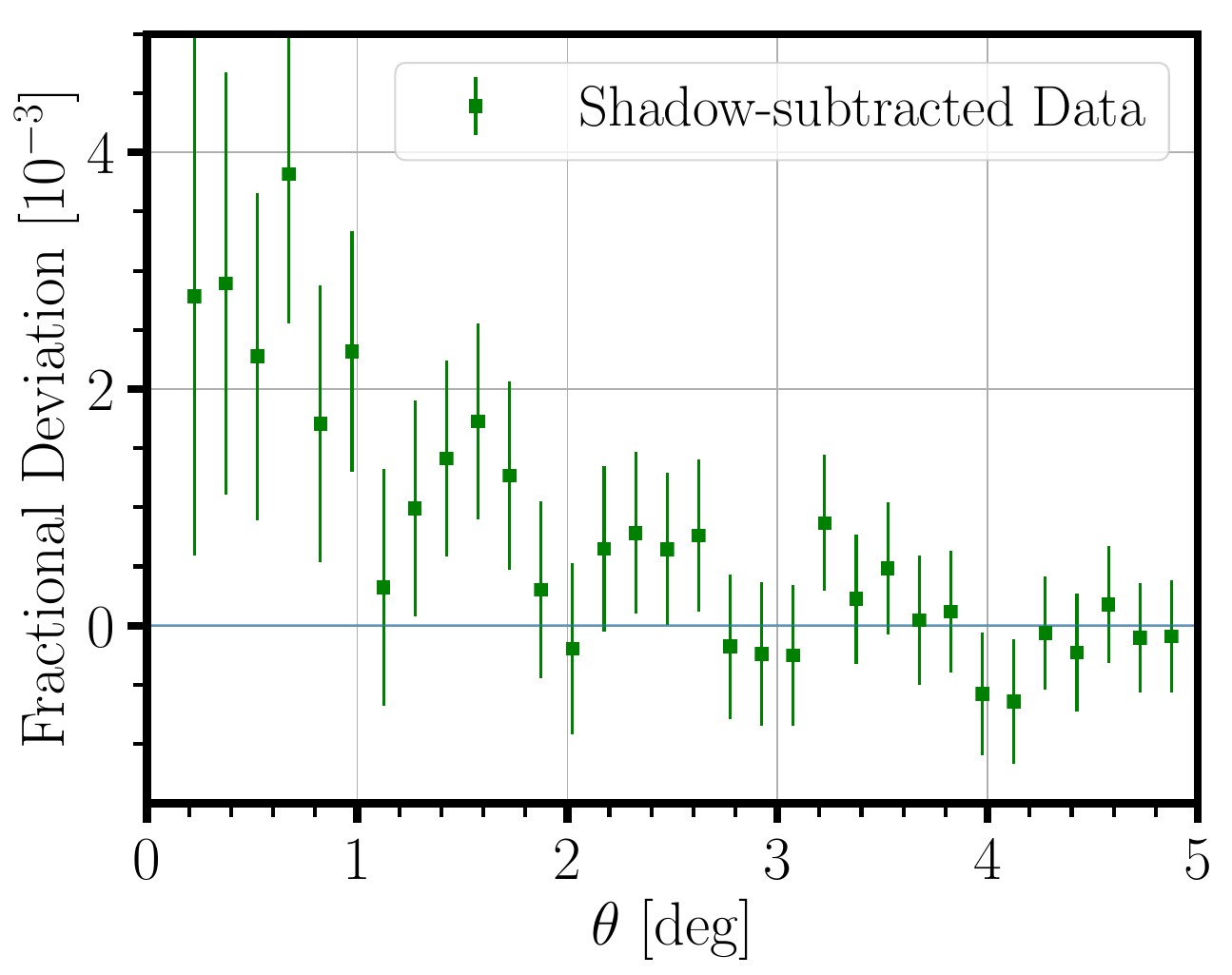}
\end{tabular}
}
\caption{Results for the combination of bins B2, B3, and B4, but otherwise as in Fig.~\ref{fig:maps}.  \textbf{Left:} The full 6.1-year data.  \textbf{Middle:} Solar-maximum period. \textbf{Right:} Solar-minimum period.}
\label{fig:minmax}
\end{figure*}

Figure~\ref{fig:minmax} shows the maps and angular profiles for the full 6.1 years of data (left), solar maximum (middle), and solar minimum (right).  Here we combine the bins B2, B3, and B4. We find a strong indication of time variation.  For the solar-maximum data, we detect only a weak signal ($3.3\sigma$), but for the solar-minimum data, we detect a strong signal ($5.9\sigma$).  We also find that the flux during the solar minimum is higher than the 6.1-year average (calculated below).  Qualitatively, these results match the time variation seen in Fermi-LAT data~\cite{Ng:2015gya, Linden:2018exo, Tang:2018wqp}.  The fact that the flux is anti-correlated with solar activity over energies 0.1 GeV to $\sim 1$ TeV, without an obvious energy dependence, is an important clue for theoretical modeling.
%%%%%%%%%%%%%%%%%%%%%%%%%%%%FIGURE2%%%%%%%%%%%%%%%%%%%%%%%%%%%%%%%%%
%%%%%%%%%%%%%%%%%%%%%%%%%%%%END FIGURE%%%%%%%%%%%%%%%%%%%%%%%%%%%%%%
%%%%%%%%%%%%%%%%%%%%%%%%%%%%%Section5%%%%%%%%%%%%%%%%%%%%%%%%%%%%%%%%%%%
%%%%%%%%%%%%%%%%%%%%%%%%%%%%%%%%%%%%%%%%%%%%%%%%%%%%%%%%%%%%%%%%%%%%%%%%

\emph{\textbf{Spectrum Slope.---}}
Although the energy range for our analysis is not wide, we can still measure the spectrum slope using the shadow-subtracted data. Here we fit for more parameters than just the flux, which means that each parameter is measured less well.

We use a forward-folded maximum-likelihood approach to obtain the flux of gamma rays from the shadow-subtracted data. Using the HAWC plug-in to the Multi-Mission Maximum Likelihood framework~\cite{Vianello:2015wwa,  2015arXiv150807479Y, Abeysekara:2021DF}, we fit for a source described by a disk of variable radius $r$ and a spectrum given by
\begin{equation}
\frac{dN}{dE} = A \left(\frac{E}{E_0}\right)^{-\gamma},
\end{equation}
where $A$ is the differential flux at the reference energy $E_0$ (1 TeV) and $\gamma$ is the spectral index.

The log-likelihood function $\mathcal{L}(A,\gamma,r)$ encodes the Poisson probability of observing $D_p$ events in each pixel $p$, given a source flux model that depends on the parameters $A$, $\gamma$, and $r$. It is written as
\begin{multline}
  \mathcal{L}(A,\gamma,r) = \sum\limits_{b = 2}^{8}\sum\limits_{p=1}^{N} \log \left\{\frac{[B_p+S_p(A,\gamma,r)]^{D_p}}{D_p!}\right\} - \\ [B_p+S_p(A,\gamma,r)],
 \label{eq:ll}  
\end{multline}
where $B_p$ is the expected number of background events in the spatial pixel $p$ and $S_p$ is the number of signal events under the assumed flux model. For the spectrum fit, we use all $N = 5940$ pixels within 5$^\circ$ of the Sun.  To obtain the best-fit parameters, we maximize the likelihood in Eq.~(\ref{eq:ll}) with respect to $A$, $\gamma$, and $r$.

We define our test statistic (TS) as
\begin{equation}
\rm TS = 2(\mathcal{L}(\hat{A},\hat{\gamma},\hat{r}) - \mathcal{L_{\rm Bkg}}), 
\end{equation}
where $\mathcal{L_{\rm Bkg}}$ is the log-likelihood for the null hypothesis (the background-only scenario) and $\hat{A}$, $\hat{\gamma}$, and $\hat{r}$ are the best-fit values of the model parameters.

Table~\ref{tab:fits} shows the results for the full 6.1 years of data, the solar-maximum period, and the solar-minimum period.  When restricting to these shorter periods, we fix the disk radius to $0.24^\circ$, which is the best-fit value for the full dataset (and close to the true value of $0.26^\circ$). Given the angular resolution in this energy range ($\sim 1^\circ$), the Sun is effectively a point source. However, a disk-like hypothesis shows a slightly higher TS than a simple point source when performing the spectral fit. While the flux at solar minimum flux is higher than that at solar maximum, the spectral slopes are consistent with each other and that for the full dataset.  The fitted values of the slope are significantly steeper than that of the cosmic rays. The spectral fits are subject to the systematic errors that result from uncertainties in the modeling of the detector response to air showers. The sources of these uncertainties are discussed in detail in Ref. \cite{HAWC:2019xhp}. In this analysis, they impact the measured flux by $\sim 15\%$. 

\begin{figure*}[h]
\includegraphics[width = 0.80\textwidth]{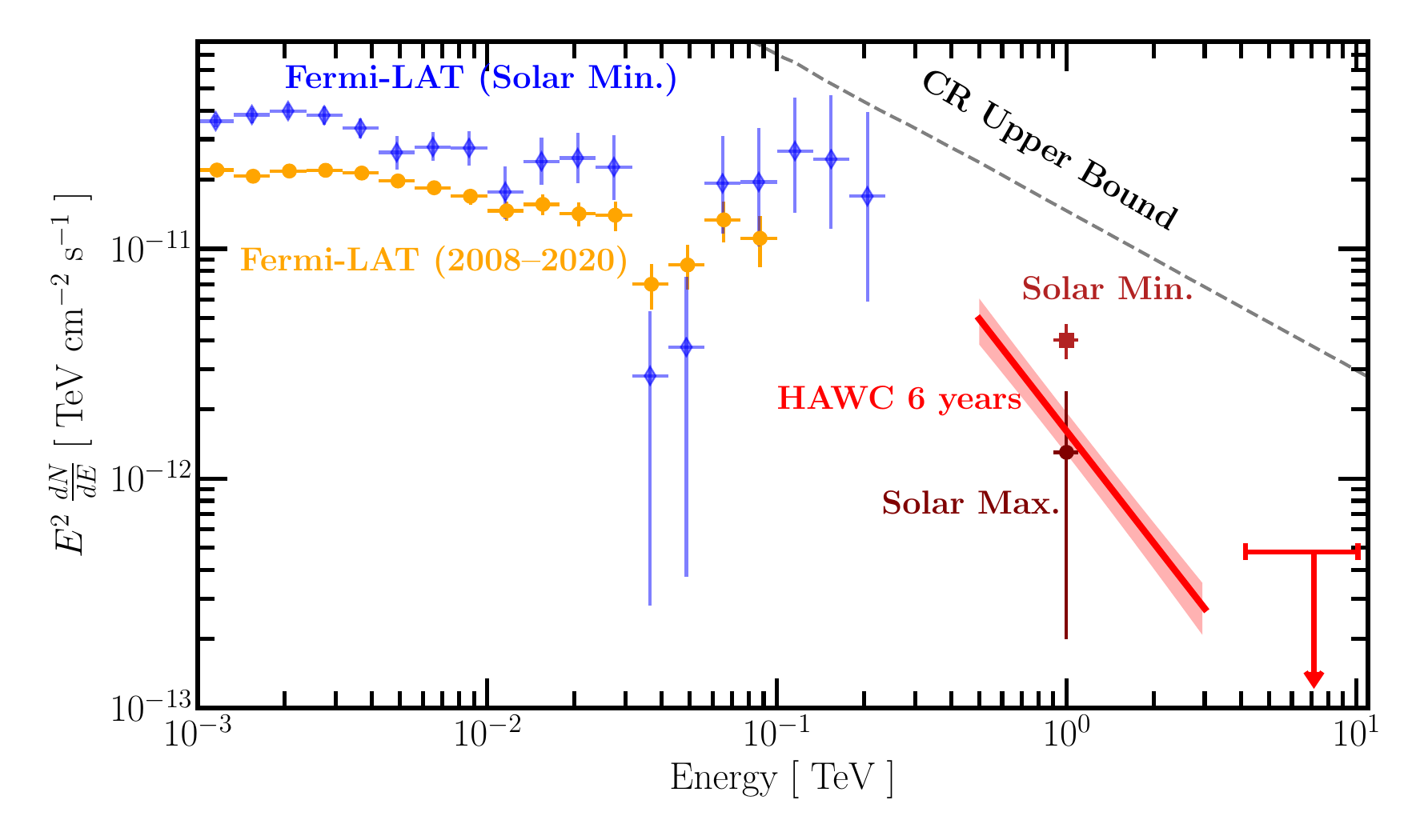}
\caption{Spectrum of the solar disk.  The 6.1-year spectrum by HAWC is shown by the red solid line and shaded band, where the latter indicates the statistical uncertainties. The 90\% CL upper limit at 7 TeV is indicated with the red arrow. The results at solar maximum and minimum are indicated just by their fluxes, placed at 1 TeV, as labeled.  The Fermi-LAT spectra over the full solar cycle ~\cite{Linden:2020lvz} and at the solar minimum \cite{Tang:2018wqp} are also shown. The dashed line shows the theoretical maximum on the gamma-ray spectrum~\cite{Linden:2018exo}.}
\label{fig:spectrum}
\end{figure*}

Figure~\ref{fig:spectrum} shows the gamma-ray spectrum of the Sun obtained with HAWC data.  We also show 1--100 GeV Fermi-LAT data over a full solar cycle (August 2008 to February 2020)~\cite{Linden:2020lvz}.  Although there is a gap between the energy ranges of Fermi-LAT and HAWC, the comparison of their fluxes, plus the steeper slope for the HAWC data, suggests a break energy at $\sim 400$ GeV, which is another important clue for theoretical modeling.

%%%%%%%%%%%%%%%%%%%%%%%%%%%%Table 1%%%%%%%%%%%%%%%%%%%%%%%%%%%%%%
\begin{table}[t!]
\begin{tabular}{c|c|c|c|c}
Data & $A \times 10^{-12}$ & $\gamma$ & $r$ [deg.] & TS\\
     &  [TeV$^{-1}$cm$^{-2}$s$^{-1}$]  & & &\\
\hline
6.1 yr      & $1.6 \pm 0.3$   & $3.62 \pm 0.14$ & $0.24 \pm 0.1$ & 45\\
Sol.\ Max.\ & $1.3 \pm 1.1$  & $3.9 \pm 0.4$ & $0.24$ (fixed) & 8.8 \\
Sol.\ Min.\ & $4.0 \pm 0.7$  & $3.52 \pm 0.14$ & $0.24$ (fixed) & 33.1
\end{tabular}
\caption{The best-fit parameters and TS values for each of the three time periods analyzed. The reported uncertainties are statistical.}
\label{tab:fits}
\end{table}
%%%%%%%%%%%%%%%%%%%%%%%%%%%%End Table%%%%%%%%%%%%%%%%%%%%%%%%%%%%%%

%%%%%%%%%%%%%%%%%%%%%%%%%%%%FIGURE 3%%%%%%%%%%%%%%%%%%%%%%%%%%%%%%
%%%%%%%%%%%%%%%%%%%%%%%%%%%%END FIGURE%%%%%%%%%%%%%%%%%%%%%%%%%%%%%

%%%%%%%%%%%%%%%%%%%%%%%%%%%%%%%%SECTION6%%%%%%%%%%%%%%%%%%%%%%%%%%%%%%%%
%%%%%%%%%%%%%%%%%%%%%%%%%%%%%%%%%%%%%%%%%%%%%%%%%%%%%%%%%%%%%%%%%%%%%%%%

\emph{\textbf{Conclusions.---}}
Probing the Sun at the highest energies is key to understanding the propagation of cosmic rays in the heliosphere, and in particular to solving the puzzles of the unexpectedly bright GeV gamma-ray emission seen by Fermi. Our TeV observations with HAWC show that the Sun continues to be an anomalously bright gamma-ray source at very high energies. The observations can be compared to the maximum possible flux assuming all the cosmic rays impinging on the solar surface are reversed and undergo hadronic interactions to produce gamma rays \cite{Linden:2018exo}. In fact, the observed flux during solar minimum is $\sim 20\%$ of the theoretical maximum emission due to cosmic-ray interactions \cite{Linden:2018exo}, indicating a remarkable efficiency of the underlying mechanism. Moreover, the observed flux is almost two orders of magnitude higher than the flux expected from the solar limb alone~\cite{Zhou:2016ljf}, indicating the important role of magnetic fields in modulating and enhancing the flux. 

Our results provide new clues about the emission mechanism.  The steeper spectral index than found for Fermi observations indicates a change in the processes, as well as a break energy $\sim 400$ GeV between the two datasets.  The measured spectrum of the Sun in HAWC data extends to an estimated maximum energy of 2.6 TeV (see S.M.). The corresponding cosmic-ray energy of $\sim 26$ TeV sets a new empirical energy scale up to which cosmic rays penetrate the photosphere and produce gamma rays under the influence of magnetic fields.  %For the typical solar magnetic field strength of $\sim 1$ G, the Larmor radius of a 26-TeV proton is $8.7 \times 10^5$ km, which is a little over 1$R_{\odot}$ (solar radius). 

Models of cosmic ray interactions in the Sun such as Refs.~\cite{Seckel:1991ffa, Mazziotta:2020uey, Li:2020gch} already under-predict the observed gamma-ray flux from the Sun in the GeV range. Our observations highlight the need for a revised framework that can explain the anomalous excess of gamma rays from the Sun also in the TeV range.
%%%%%%%%%%%%%%%%%%%%%%%%%%%%%%%%%%%%%%%%%%%%%%%%%%%%%%%%%%%%%%%%%%%%%%%%
%%%%%%%%%%%%%%%%%%%%%%%%%%%%%%%%%%%%%%%%%%%%%%%%%%%%%%%%%%%%%%%%%%%%%%%%

%TC:ignore

\bigskip
\begin{acknowledgements}

{\bf\emph{Acknowledgments.---}}

We thank Jung-Tsung Li for useful comments on the manuscript. We acknowledge the support from: the US National Science Foundation (NSF); the US Department of Energy Office of High-Energy Physics; the Laboratory Directed Research and Development (LDRD) program of Los Alamos National Laboratory; Consejo Nacional de Ciencia y Tecnolog\'ia (CONACyT), M\'exico, grants 271051, 232656, 260378, 179588, 254964, 258865, 243290, 132197, A1-S-46288, A1-S-22784, c\'atedras 873, 1563, 341, 323, Red HAWC, M\'exico; DGAPA-UNAM grants IG101320, IN111716-3, IN111419, IA102019, IN110621, IN110521, IN102223; VIEP-BUAP; PIFI 2012, 2013, PROFOCIE 2014, 2015; the University of Wisconsin Alumni Research Foundation; the Institute of Geophysics, Planetary Physics, and Signatures at Los Alamos National Laboratory; Polish Science Centre grant, DEC-2017/27/B/ST9/02272; Coordinaci\'on de la Investigaci\'on Cient\'ifica de la Universidad Michoacana; Royal Society - Newton Advanced Fellowship 180385; Generalitat Valenciana, grant CIDEGENT/2018/034; The Program Management Unit for Human Resources \& Institutional Development, Research and Innovation, NXPO (grant number B16F630069);; Coordinaci\'on General Acad\'emica e Innovaci\'on (CGAI-UdeG), PRODEP-SEP UDG-CA-499; Institute of Cosmic Ray Research (ICRR), University of Tokyo, H.F. acknowledges support by NASA under award number 80GSFC21M0002. We also acknowledge the significant contributions over many years of Stefan Westerhoff, Gaurang Yodh, and Arnulfo Zepeda Dominguez, all deceased members of the HAWC collaboration. Thanks to Scott Delay, Luciano D\'iaz and Eduardo Murrieta for technical support.

The work of J.F.B.\ was supported by NSF grant No.\ PHY-2012955. The work of B.Z. was supported by the Simons Foundation. The work of A.H.G.P, B.Z., K.N., J.F.B., M.N. and T.L. was supported in part by NASA Grant No.\ 80NSSC20K1354. The work of A.H.G.P. and J.F.B. was additionally supported in part by NASA Grant No.\ 80NSSC22K0040. The work of TL was supported by the ERC under grant 742104, the SNSA under contract 117/19 and VR under contract 2019-05135. The work of KCYN is supported by the RGC of HKSAR, Project No. 24302721.

\end{acknowledgements}

%%%%%%%%%%%%%%%%%%%%%%%%%%%%%%%%%%%%%%%%%%%%%%%%%%%%%%%%%%%%%%%%%%%%%%%%
%%%%%%%%%%%%%%%%%%%%%%%%%%%%%%%%%%%%%%%%%%%%%%%%%%%%%%%%%%%%%%%%%%%%%%%%

%\clearpage

\bibliography{mbib}

%%%%%%%%%%%%%%%%%%%%%%%%%%%%%%%%%%%%%%%%%%%%%%%%%%%%%%%%%%%%%%%%%%%%%%%%
%%%%%%%%%%%%%%%%%%%%%%%%%%%%%%%%%%%%%%%%%%%%%%%%%%%%%%%%%%%%%%%%%%%%%%%%

\clearpage

\onecolumngrid
\appendix

\ifx \standalonesupplemental\undefined
\setcounter{page}{1}
\setcounter{figure}{0}
\setcounter{table}{0}
\setcounter{equation}{0}
\setcounter{secnumdepth}{2}
\fi

\renewcommand{\thepage}{Supplemental Material -- S\arabic{page}}
\renewcommand{\theequation}{S\arabic{equation}}
\renewcommand{\figurename}{SUPPL. FIG.}
\renewcommand{\tablename}{SUPPL. TABLE}

\clearpage
\newpage

\centerline{\Large {\bf Supplemental Material for}}
\medskip

\centerline{\Large \emph{The TeV Sun Rises: Discovery of Gamma Rays from the Quiescent Sun with HAWC}}
\medskip

\centerline{\large HAWC Collaboration with J. F. Beacom, T. Linden, K. C. Y. Ng, A. H. G. Peter and B. Zhou}
\bigskip

We provide additional details in the appendices below to aid the reading of the paper. Appendix \ref{app:hawc} provides the estimated HAWC sensitivity, and the median energies and angular resolutions for the bins used in this work. Appendix \ref{app:shadow} gives details regarding the shadow measurement and subtraction. Appendices \ref{app:offsun}, \ref{app:moon} and \ref{app:counts} describe the various cross-checks performed to verify the robustness of the results presented in this paper. Finally, \ref{app:bins} shows maps of the Sun from all the bins used in this work.  

\section{HAWC Sensitivity\label{app:hawc}}
\begin{figure*}[ht]
\includegraphics[width = 0.6\textwidth]{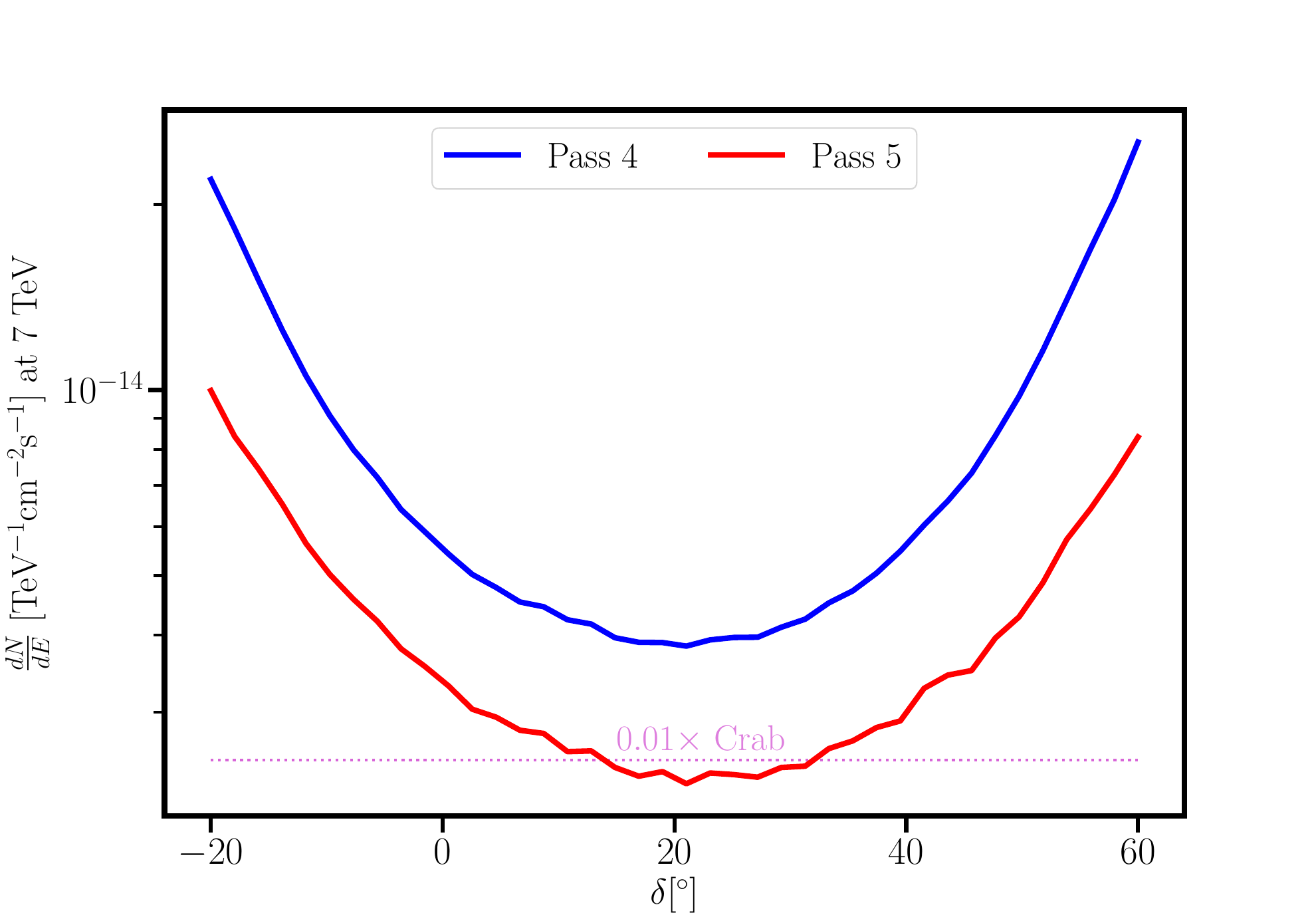}
\caption{HAWC's sensitivity to a point-source search as a function of declination. The differential flux sensitivity at a characteristic energy of 7 TeV is shown for a  spectral hypotheses of a Crab-like source \cite{HAWC:2019xhp}. The sensitivity corresponding to 2241 days of Pass 4(5) data is shown in black (red). The dotted blue line shows $1\%$ of the Crab flux.}
\label{fig:sensi}
\end{figure*}

Figure \ref{fig:sensi} shows the improvement in HAWC's differential sensitivity at 7 TeV for a source with a spectrum similar to the Crab nebula \cite{HAWC:2019xhp} between the Pass 4 and Pass 5 data samples. HAWC is able to achieve a sensitivity down to $< 1\%$ of the ``Crab flux'' for certain source declinations, and an improvement in sensitivity of $\geq 2$ relative to Pass 4 for all source declinations. With the new data sample, HAWC is able to achieve an angular resolution of $0.5^\circ$ near 1 TeV for the zenith angles over which the Sun transits above the observatory. Table \ref{tab:bins} shows the median gamma-ray energies and the central 68\% containment radius of the point-spread-function for the bins used in this analysis. We only use B2--B8, as other bins do not contain sufficient statistics for a robust analysis of the Sun. The data still cover a wide energy range from $\sim 600$ GeV to $> 10$ TeV.

%%%%%%%%%%%%%%%%%%%%%%%%%%%%Figure%%%%%%%%%%%%%%%%%%%%%%%%%%%

%%%%%%%%%%%%%%%%%%%%%%%%%%%%End Figure%%%%%%%%%%%%%%%%%%%%%%%%%%%

%%%%%%%%%%%%%%%%%%%%%%%%%%%%Table 2%%%%%%%%%%%%%%%%%%%%%%%%%%%%%%
\begin{table}[ht!]
\begin{tabular}{c|c|c}
Analysis Bin & Median Energy [TeV] & Angular Resolution [$^\circ$] \\
\hline
B2     & $0.6$  & $0.8$ \\
B3  & $1.0$ & $0.5$ \\
B4  & $1.7$ & $0.4$\\
B5  & $2.8$ & $0.3$\\
B6  & $4.6$ & $0.2$\\
B7  & $7.4$ & $0.2$\\
B8  & $11.5$ & $0.2$\\
\end{tabular}
    \caption{The median energies and the angular resolution in each analysis bin for a source located at the average solar zenith  with respect to HAWC. The angular resolution is defined as the 68\% containment radius for the angular distribution of the photons.}
    \label{tab:bins}
\end{table}
%%%%%%%%%%%%%%%%%%%%%%%%%%%%End Table%%%%%%%%%%%%%%%%%%%%%%%%%%%%%%

\section{Shadow Subtraction\label{app:shadow}}
The technique of shadow measurement and subtraction relies on the fact that hadronic cosmic rays vastly outnumber gamma rays in data by a facator of $\sim 10^4--10^6$. This allows for an accurate measurement of the spatial profile of the shadow without contamination by gamma-ray events. HAWC skymaps consist of the reconstructed directions of air-showers projected on a two dimensional grid, such that the center of the grid is at the Sun. Before gamma-hadron cuts, i.e., in the shadow maps, in a given pixel there are $N_{\rm shadow}$ data counts, while the number of background cosmic rays are $\left< N_{\rm bkg}^{\rm CR}\right >$. The deficit or excess in any pixel can be measured w.r.t. to the background. The fractional deviation $\Delta I$ from background in any pixel is then given by

\begin{equation}
    \Delta I = \frac{N_{\rm shadow} - \left< N_{\rm bkg}^{\rm CR}\right >}{\left< N_{\rm bkg}^{\rm CR}\right >}.
    \label{eq:sup1}
\end{equation}

The quantity $\Delta I$ in every pixel near the Sun is negative due to the deficit of CR in the data. It  describes our ``true'' (high-statistics) measurement of the amplitude of the shadow. Gamma-hadron cuts result in an overall reduction in the number of background and data counts in each pixel. However, the amplitude of the true shadow remains unchanged. Therefore, analogous to equation \ref{eq:sup1}, we multiply $\Delta I$ by the measured background after gamma-hadron cuts $\left< N_{\rm bkg}\right >$, to yield the number of counts that need to be subtracted from the data in each pixel to account for the shadow.

\section{Cross-checks on Off-Sun regions\label{app:offsun}}
We also validate the analysis method on background-only or off-source ``fake'' Sun regions with simulated shadows without a gamma-ray signal. We choose 52 regions at 1$^\circ$ intervals on the Sun's path in the sky. The fake Suns lie at an angular distance between 5$^\circ$ and 30$^\circ$ from the true position of the Sun. At each fake Sun position in the cosmic-ray data, we remove events to simulate a Sun shadow. The simulated shadow in a given bin follows the same spatial profile as the observed shadow. We introduce a similar shadow with Poisson fluctuations in the data after applying gamma-hadron cuts. We then repeat the shadow-subtraction procedure described in the main text to obtain the residual excess from the fake Suns. 

\begin{figure*}[ht!]
\centering
\makebox[0.4\width][c]{
\begin{tabular}{@{}cc@{}}
\includegraphics[width=0.5\textwidth]{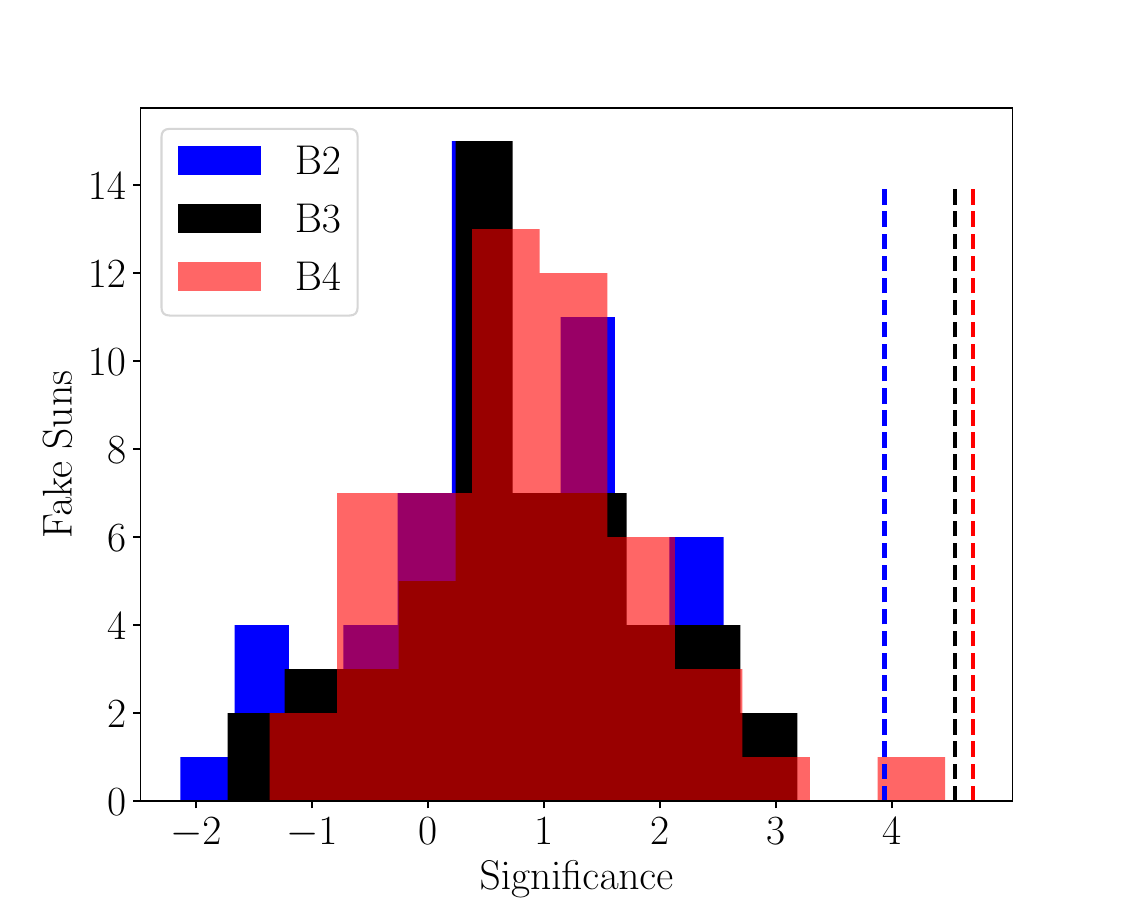} &
\includegraphics[width=0.5\textwidth]{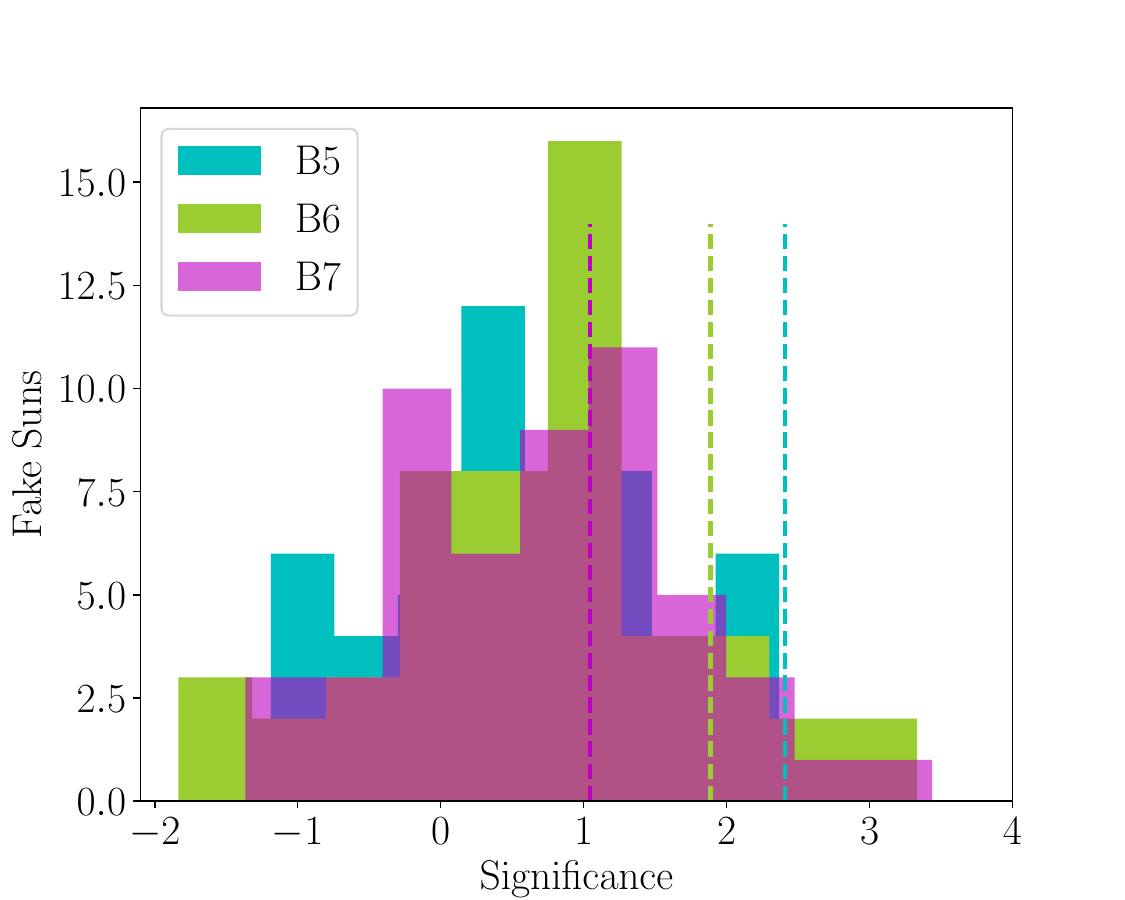} \\

\end{tabular}
}
\caption{The maximum significance of an excess observed within 0.26$^\circ$ of a fake Sun region for B2, B3 and B4 (left) and B5, B6 and B7 (right). The dashed vertical lines show the significance for the true Sun in the corresponding bin.}
\label{fig:offsun}
\end{figure*}

Figure \ref{fig:offsun} shows the distribution of the significances obtained from the fake Sun regions for each bin. It can be seen that for B2 to B4, the most significant excess around a fake Sun is always less than the significance from the true Sun. Moreover, none of the positive fluctuations from fake Suns exhibit the spatial shape of the signal from the true Sun. This test demonstrates that the observed gamma-ray excess is unique to the Sun and not an artifact of the analysis method.

The median significance of the excess from the off-Sun regions is slightly greater than zero. This is because the energy distribution of cosmic rays in the data before and after gamma-hadron separation changes by a small amount. Higher energy cosmic rays are more easily rejected than lower energy ones. Since the central position of the shadow in a given bin is determined by the energy distribution of cosmic rays in that bin, the change in energies causes a small shift in the position of the shadow before and after gamma-hadron cuts. This may result in a net positive excess upon subtraction. The median significance of the excess from the off-Sun regions is between 0 and $0.8\sigma$, showing that the impact of this systematic is negligible and does not affect our overall results.  

%-------------------------------------------------------FIGURE-----------------------------------------------------------------%

%-------------------------------------------------------END FIGURE----------------------------------------------------------%

\section{Cross-checks on the Moon\label{app:moon}}

%-------------------------------------------------------FIGURE-----------------------------------------------------------------%
\begin{figure*}[ht]
\centering
\makebox[1\width][c]{
\begin{tabular}{@{}cc@{}}
\includegraphics[width=0.47\textwidth]{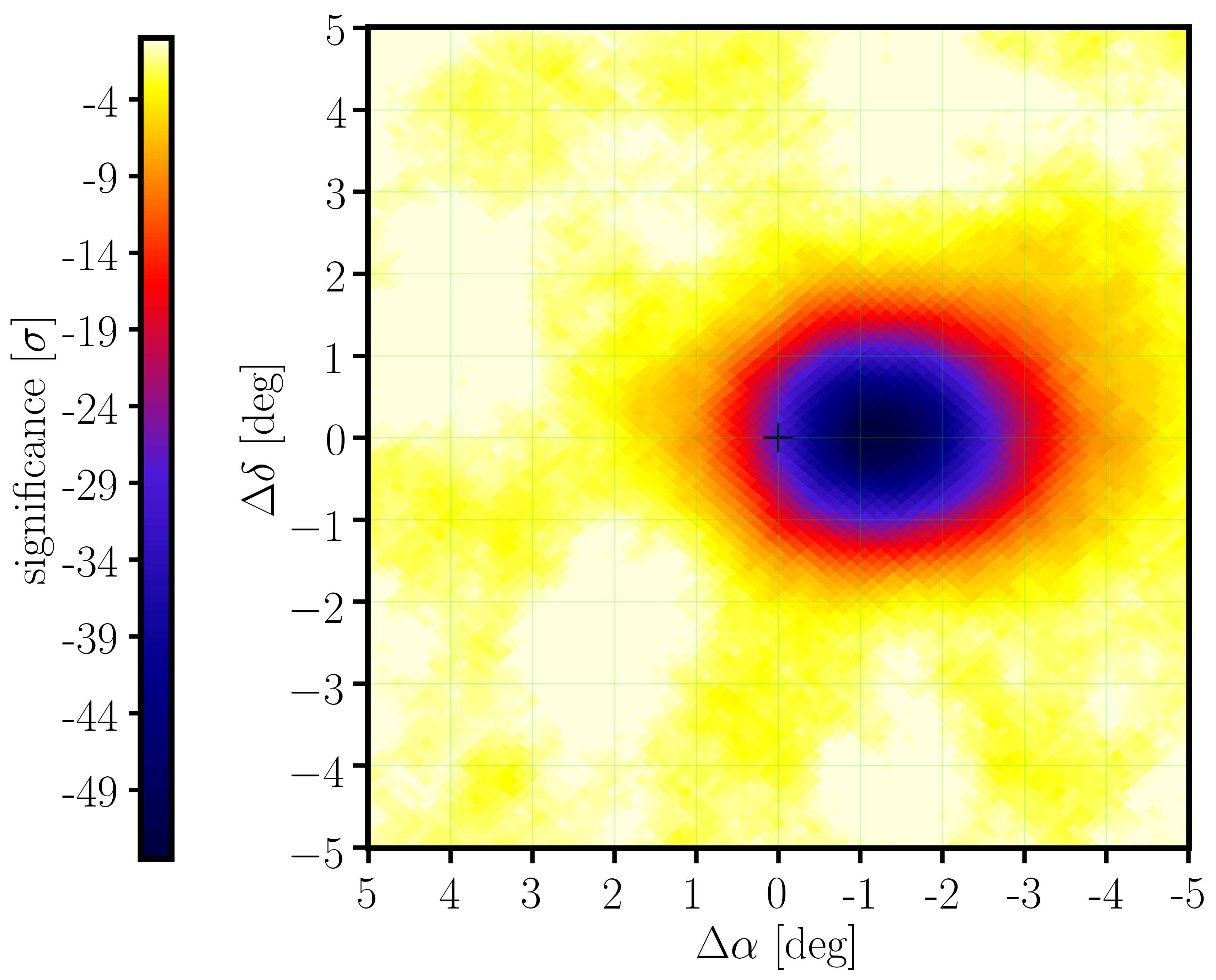} &
\includegraphics[width=0.47\textwidth]{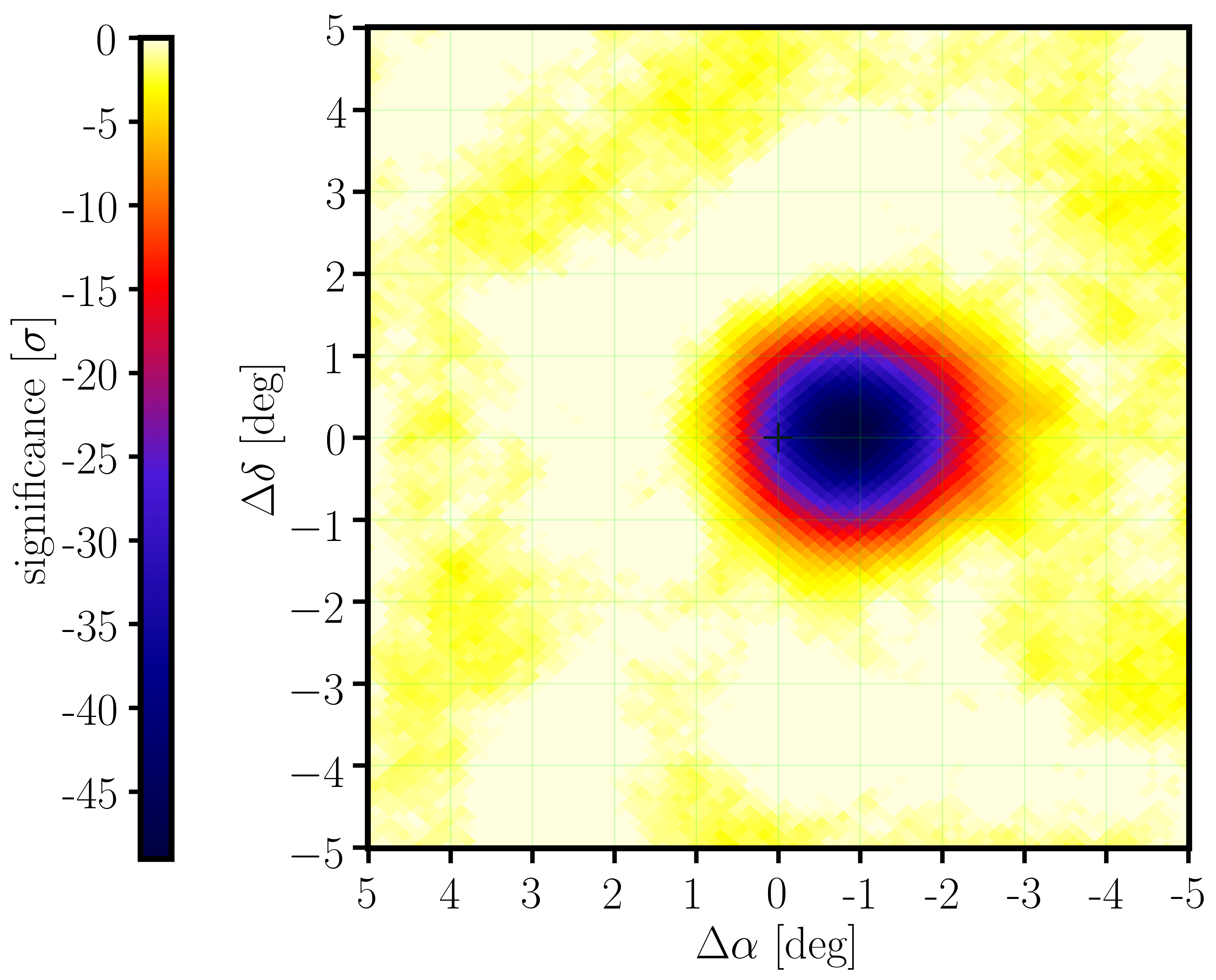} \\
\includegraphics[width=0.47\textwidth]{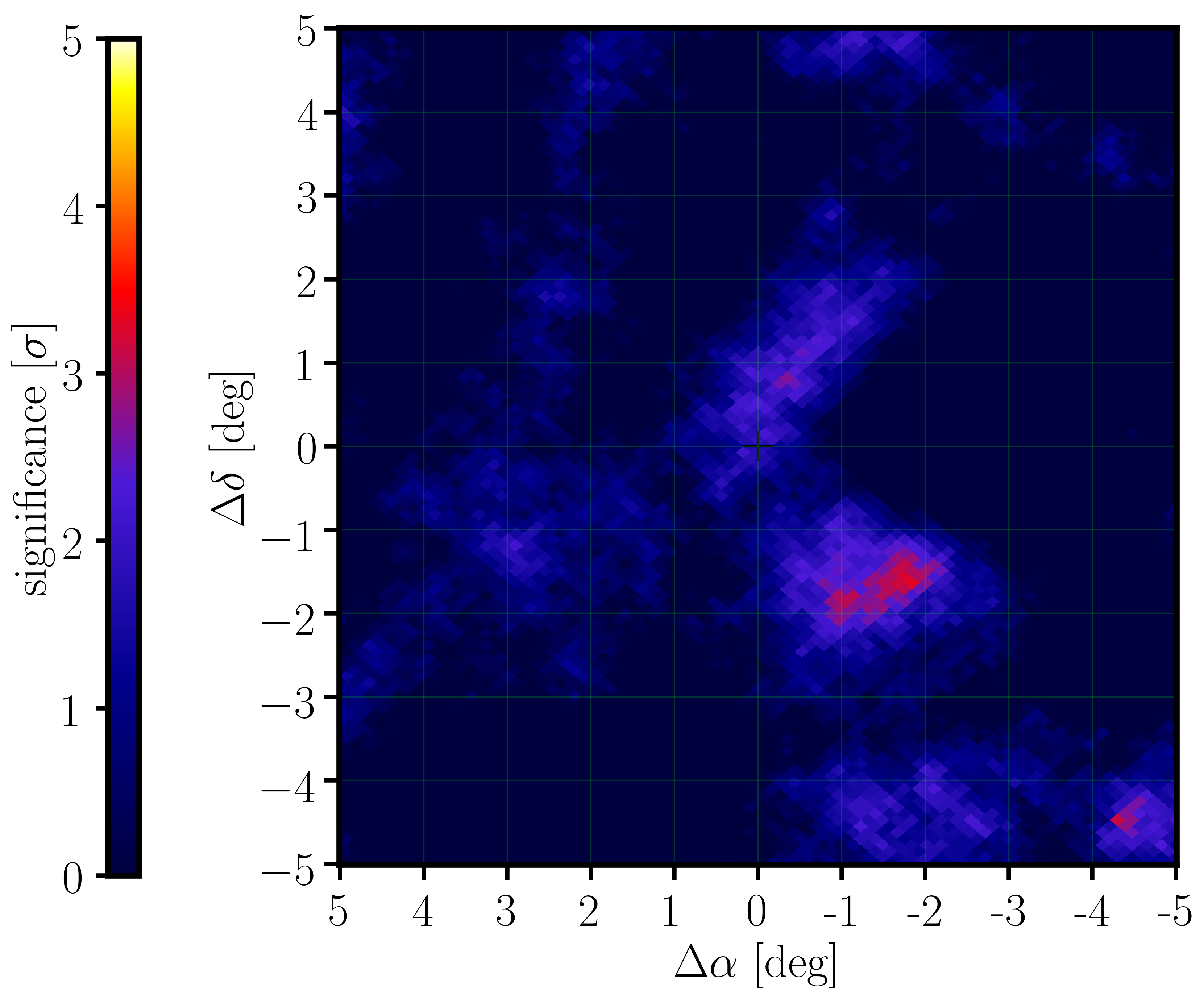} &
\includegraphics[width=0.47\textwidth]{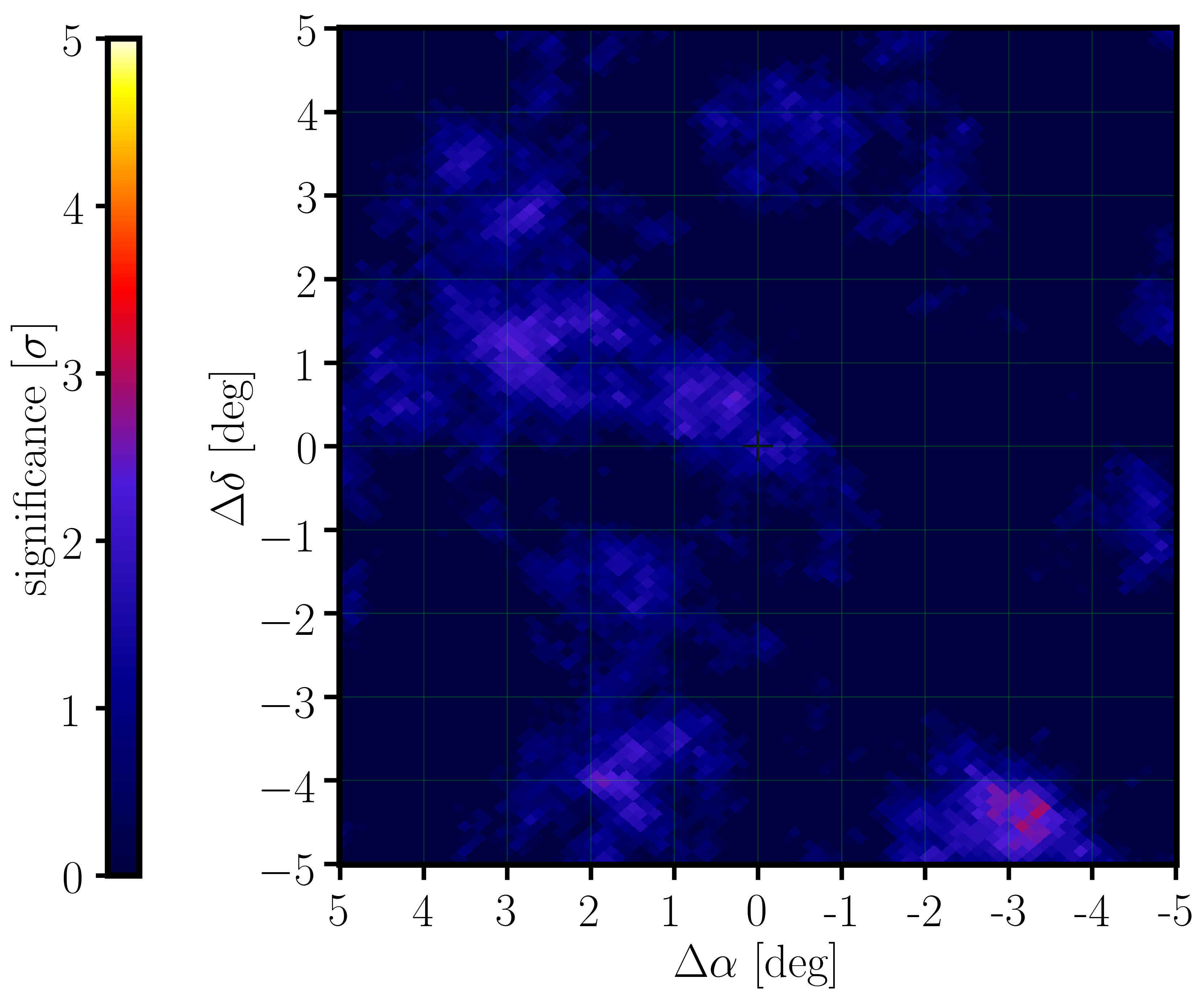} \\
\includegraphics[width=0.47\textwidth]{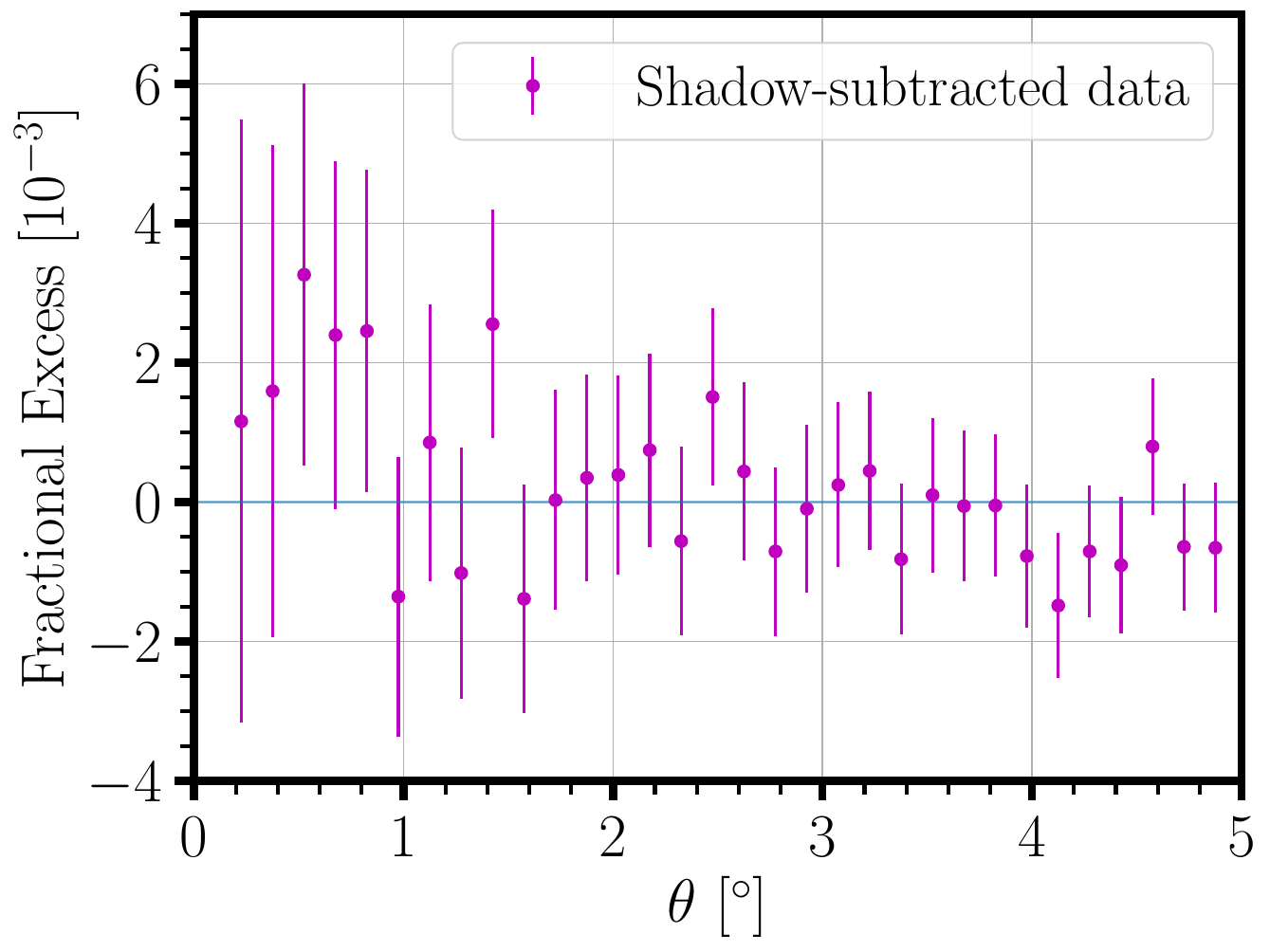}&
\includegraphics[width=0.47\textwidth]{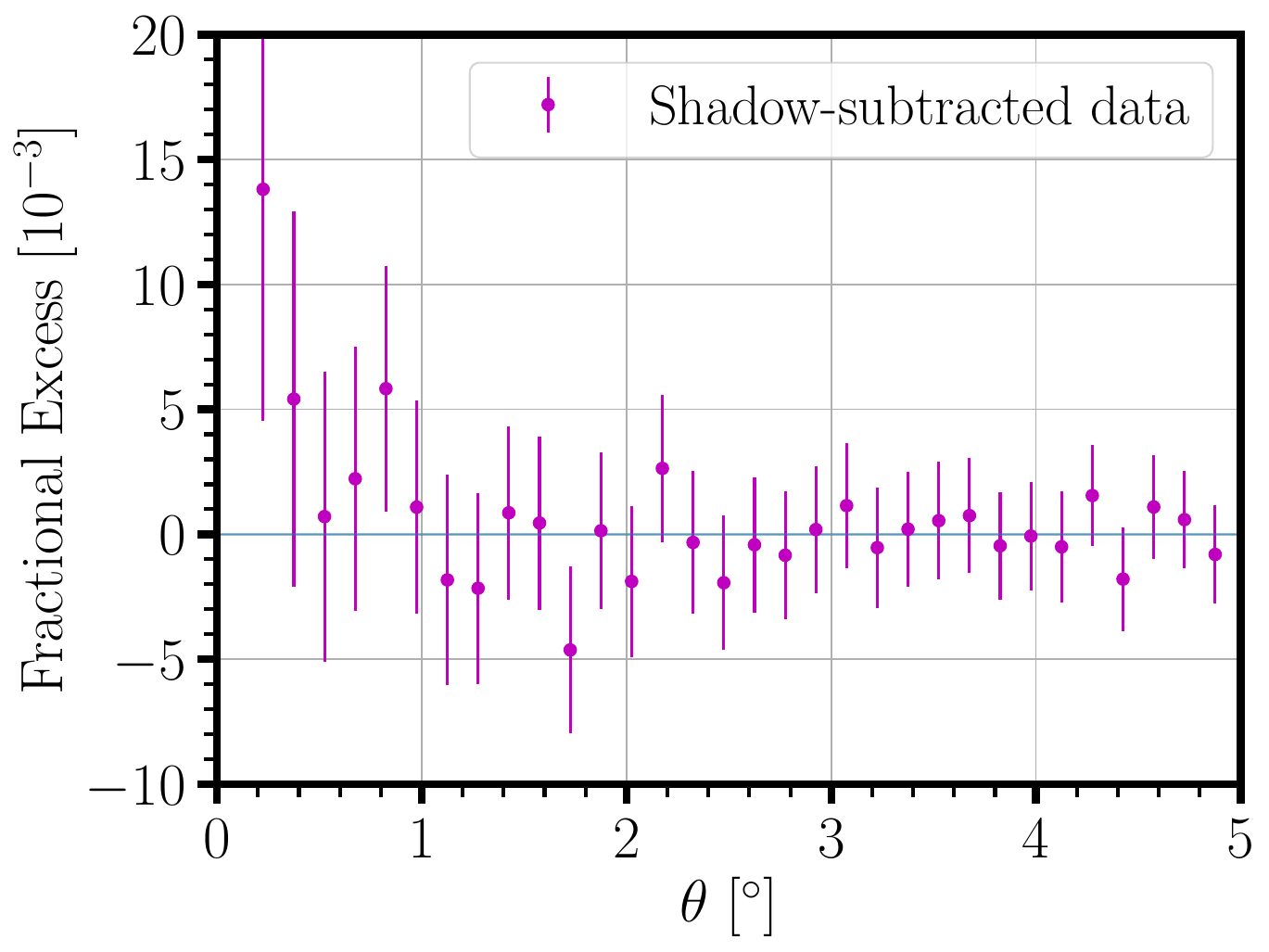}\\
\end{tabular}
}
\caption{The Moon in B3 (left) and B4 (right). The top row shows the cosmic-ray shadow of the Moon after applying the same quality cuts as applied to the Sun. The middle row shows the shadow-subtracted gamma-ray data in the same region. The Moon does not appear to be a significant source of gamma-ray emission as seen in these maps. The bottom row shows the radial excess at intervals of $0.15^\circ$.}
\label{fig:moon}
\end{figure*}

%-------------------------------------------------------END FIGURE----------------------------------------------------------%
Similar to the Sun, the Moon also casts a cosmic-ray shadow by blocking part of the incoming flux of cosmic rays \cite{Abeysekara:2018syp}. However, as opposed to the low-density solar atmosphere, the ultra-thin moon atmosphere and its hard surface implies very little hadronic gamma-ray production above a few GeV \cite{DeGaetano:2021dgh}.  Therefore, it can be used as another location to test our analysis methods. Since the Moon is not expected to be a source of TeV gamma rays, the analysis should yield a result consistent with background or off-Sun regions. We repeat the whole analysis pipeline as described in the main text using data around the Moon. Figure \ref{fig:moon} shows the results, with the observed Moon shadow in the top panel and the shadow-subtracted data in the bottom panel. It can be seen that there are only weak hints ($\sim 2\sigma$) of a positive residual excess around the moon in the two bins where the Sun is most significantly detected. The spatial profile of the positive fluctuations is not consistent with a disk-like source like the Sun. This study further validates our analysis, and strengthens the case for the observed excess around the Sun being unique.

\section{Cut-and-count method\label{app:counts}}
\begin{figure}[h!]
\includegraphics[width = 0.51\textwidth]{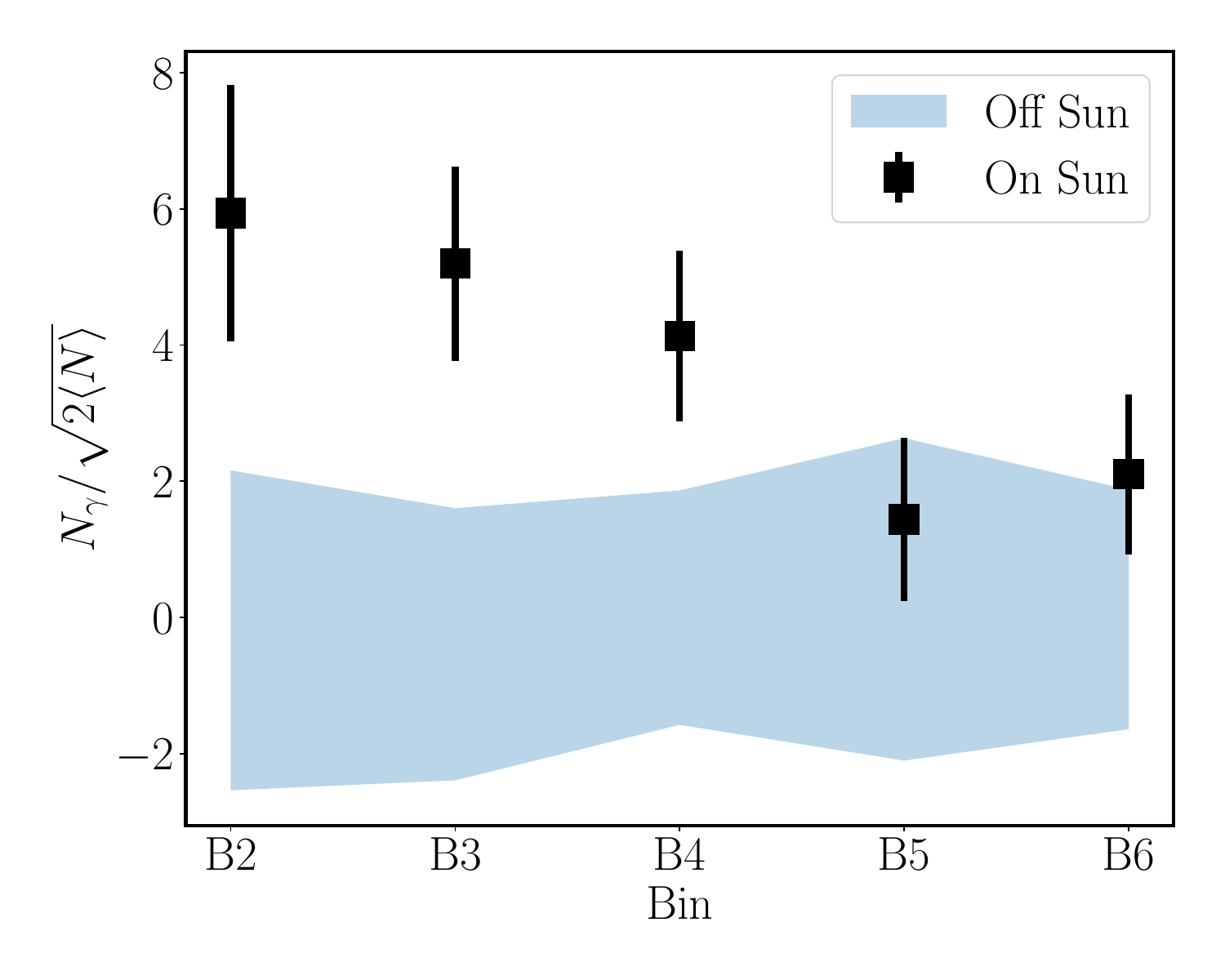}
\caption{The estimated gamma-ray excess divided by the estimated error on the background (square root of background) for on-Sun and off-Sun regions.}
\label{fig:cutcount}
\end{figure}

\noindent Finally, we can use a different method to estimate the gamma-ray excess from the Sun that does not rely on subtracting the shadow. This method was used in Ref. \cite{HAWC:2018rpf} to search for gamma rays from the Sun in three years of Pass 4 HAWC data. We show in Ref. \cite{HAWC:2018rpf} that the estimated number of net gamma-ray excess events summed over all pixels in a given region of interest (RoI) $N_{\gamma}$ is given by,

\begin{equation}
    N_{\gamma} = \frac{N_{\rm cuts} - \epsilon_{\rm CR}N_{\rm shadow}}{\epsilon_\gamma - \epsilon_{\rm CR}},
\end{equation}

where $N_{\rm shadow}$ and $N_{\rm cuts}$ are the total  number of observed events (both gammas and hadrons) above background in the shadow data and post-cuts data respectively; and  $\epsilon_\gamma$ and $\epsilon_{\rm CR}$ are the respective fractions of gamma-ray events and cosmic-ray events remaining after applying gamma-hadron cuts. These fractions are obtained from Monte Carlo simulations. The net gamma-ray excess relative to the approximate uncertainty in the background events  $\left< N_{\rm bkg}\right >$ is shown in Figure \ref{fig:cutcount}. A clear excess above the off-Sun regions is seen in the first three bins which is consistent with the main analysis presented in this work.

%%%%%%%%%%%%%%%%%%%%%%%%%%%%Figure%%%%%%%%%%%%%%%%%%%%%%%%%%%

%%%%%%%%%%%%%%%%%%%%%%%%%%%%End Figure%%%%%%%%%%%%%%%%%%%%%%%%%%%
%-------------------------------------------------------FIGURE-----------------------------------------------------------------%

\begin{figure*}[tp!]
\centering
\makebox[0.4\width][c]{
\begin{tabular}{@{}ccc@{}}
\includegraphics[width=0.37\textwidth]{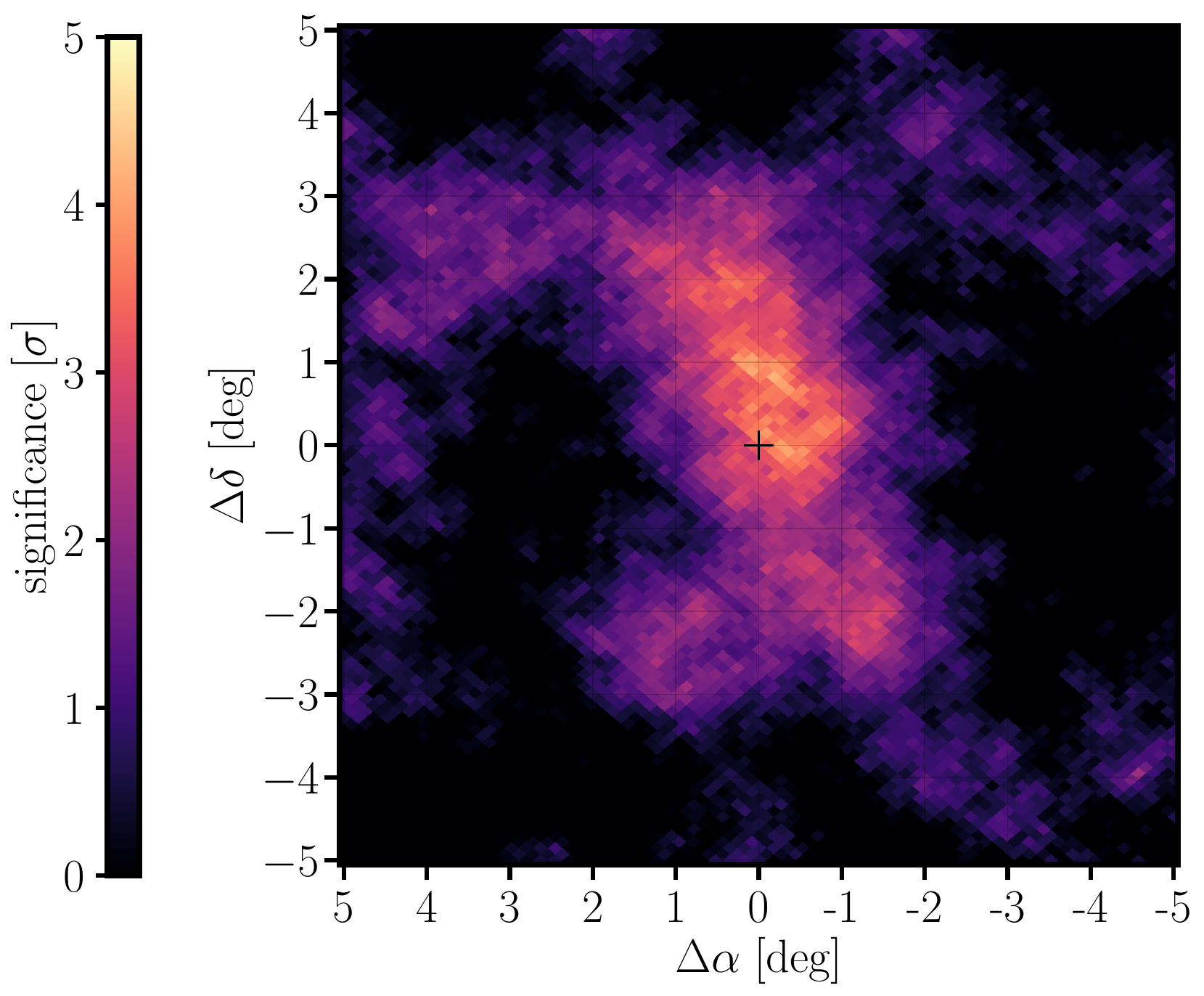} &
\includegraphics[width=0.37\textwidth]{2dmap_binB3C0_residual_gamma.pdf} &
\includegraphics[width=0.37\textwidth]{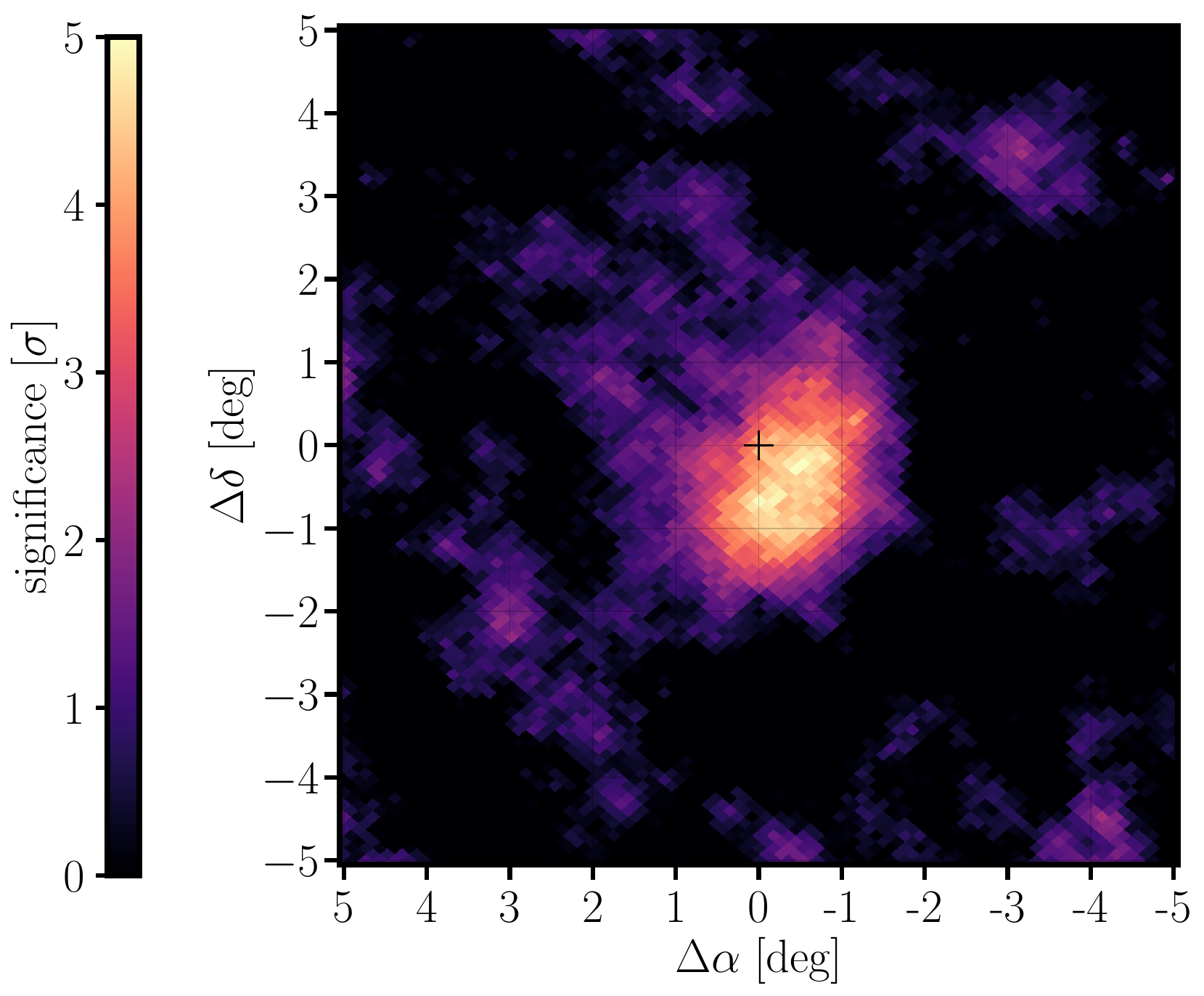} \\
\includegraphics[width=0.37\textwidth]{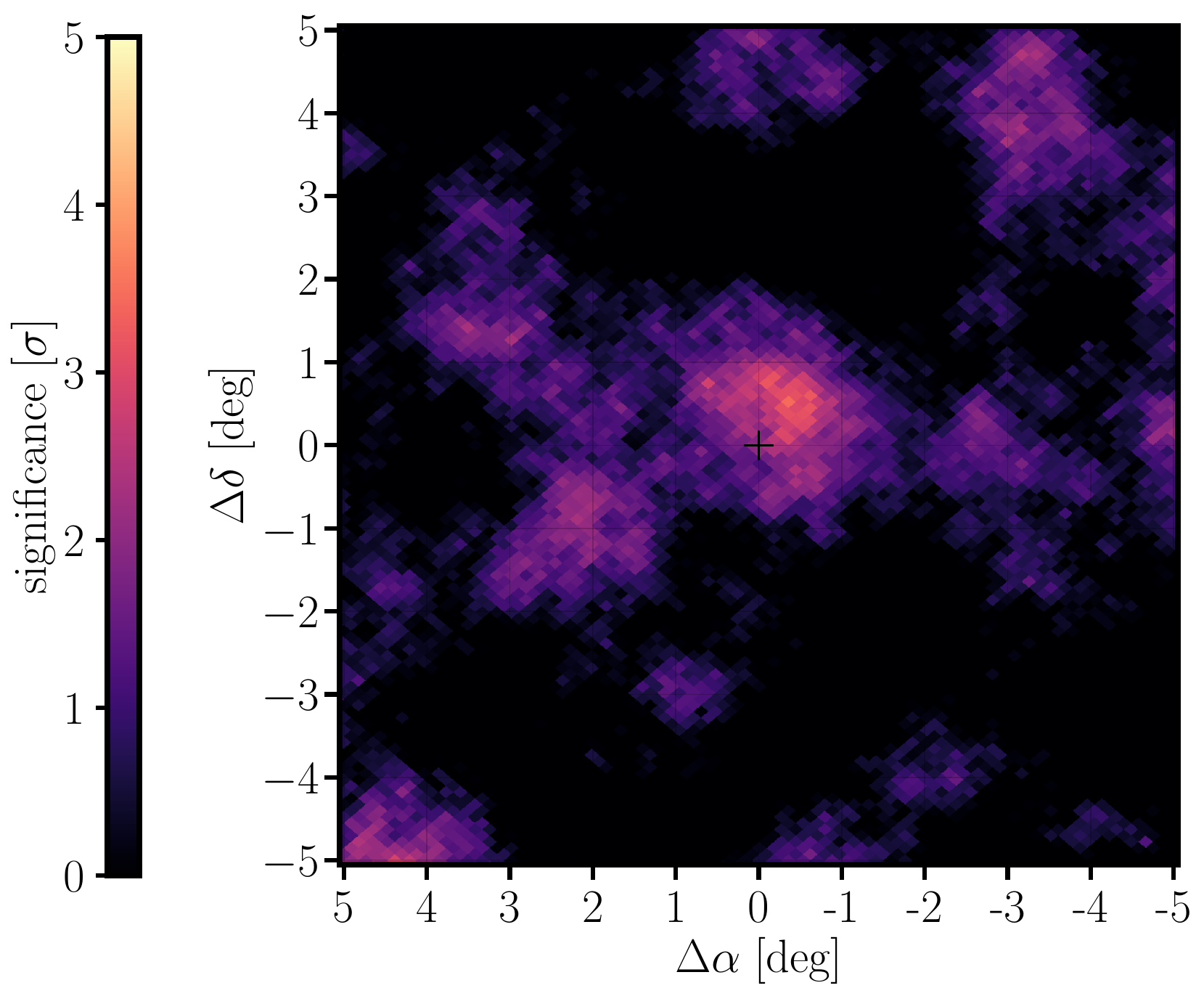} &
\includegraphics[width=0.37\textwidth]{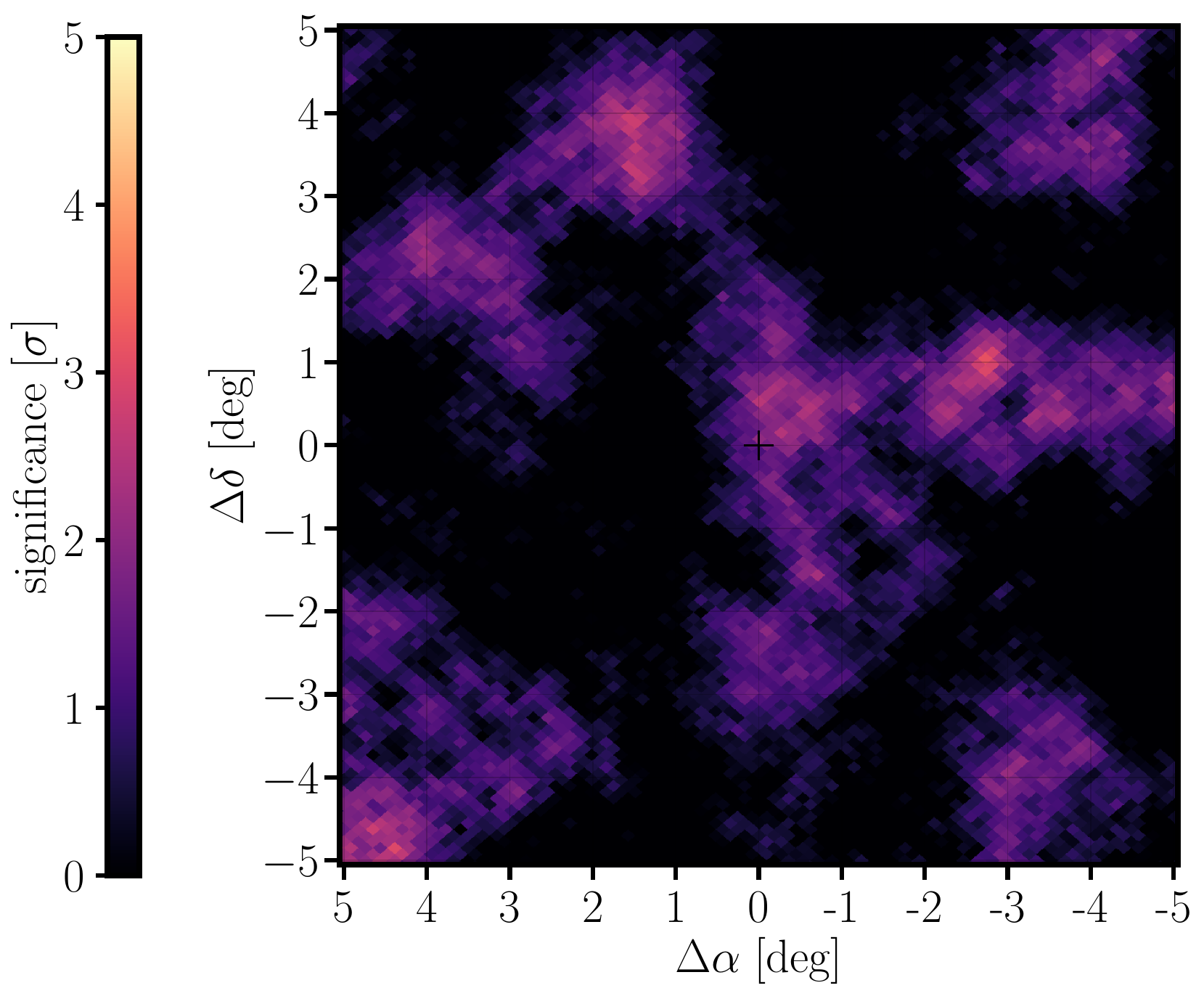}&
 \includegraphics[width=0.37\textwidth]{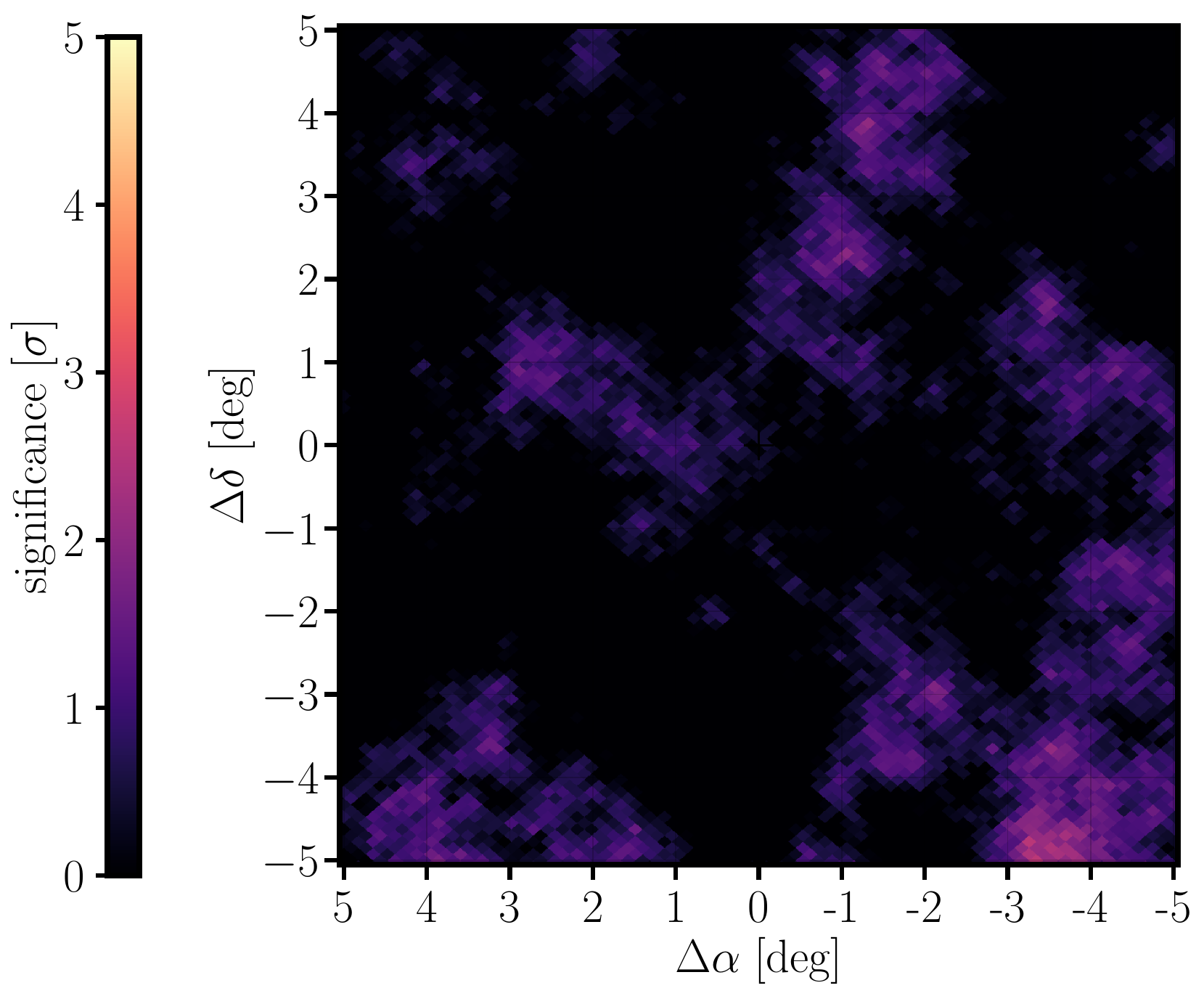}\\
\end{tabular}
}
\caption{\textbf{Top:} The shadow-subtracted significance maps for B2, B3 and B4. \textbf{Bottom:} Maps for B5, B6 and B7. The Sun is detected at $>4\sigma$ in all the lower energy bins in the top row.}
\label{fig:all_bins}
\end{figure*}

%-------------------------------------------------------END FIGURE----------------------------------------------------------%

\section{Maps of Additional Bins and Highest Energy Estimate \label{app:bins}}

Figure \ref{fig:all_bins} shows plots from bins that were not shown in the main text. The spectral fit made use of data from bins B2--B8.

\subsection{Highest Energy}
We cannot perform an event-by-event energy estimate in this analysis. We can obtain an approximate estimate of the highest energy gamma-ray event from the Sun by looking for a ``hard cutoff method'' in the measured spectrum. We fit the data again to a new function which is the product of the best-fit spectrum and a step function, $f(E)$: 

\begin{equation}
A \left(\frac{E}{E_0}\right)^{-\gamma} \cdot f(E) = 
\begin{cases} 
      0 & E < 0. \\
      1 & \rm lower bound \leq $E$ \leq \rm upper bound. \\
      0 &  E > \rm upper bound.
\end{cases}
\end{equation}

The only free parameter of the fit is the value of the upper (lower) bound for determining the highest (lowest) energy. The TS is defined as the ratio of the minimized likelihood with and without the step function. We take the value of the energy bound at $\Delta$ TS $= $1 as the maximum estimated energy at 1$\sigma$. For the Sun, this value is 2.6 TeV.

%TC:endignore
%%%%%%%%%%%%%%%%%%%%%%%%%%%%%%%%%%%%%%%%%%%%%%%%%%%%%%%%%%%%%%%%%%%%%%%%
%%%%%%%%%%%%%%%%%%%%%%%%%%%%%%%%%%%%%%%%%%%%%%%%%%%%%%%%%%%%%%%%%%%%%%%%

\end{document}